\documentclass[a4paper,12pt]{article}

\usepackage[utf8]{inputenc}
\usepackage{graphicx}
\usepackage{verbatim}
\usepackage{amsmath}
\usepackage{amsfonts}
\usepackage{amssymb}
\usepackage{color}
\usepackage{latexsym}
\usepackage{caption}
\usepackage{subcaption}

\usepackage{mathtools}
\usepackage{stmaryrd}

\usepackage[authoryear]{natbib}
\usepackage{url}
\usepackage{lineno}

\newcommand{\beq}{\begin{equation}}
\newcommand{\eeq}{\end{equation}}
\newcommand{\beqs}{\begin{equation*}}
\newcommand{\eeqs}{\end{equation*}}

\newcommand{\ev}{\mathrm{E}}

\newcommand{\commblue}[1]{#1}

\newcommand{\sra}{\shortrightarrow}
\begin{document}
\title{The expected neutral frequency spectrum of linked sites}
\author{Luca Ferretti$^{1,2,3}$\footnote{Email: luca.ferretti@gmail.com}, Alexander Klassmann$^{4}$, Emanuele Raineri$^{5}$, \\
Thomas Wiehe$^{4}$, 
Sebastian E. Ramos-Onsins$^{6}$, Guillaume Achaz$^{2,3}$}
\date{}
\maketitle

\small
(1) The Pirbright Institute, United
Kingdom. (2) Atelier de Bioinformatique, ISyEB (UMR 7205 CNRS-MNHN-UPMC-EPHE), Paris, France (3) Stochastic Models for the
Inference of Life Evolution, CIRB (UMR 7241 CNRS-INSERM), Collège de France, Paris. (3)  (4) Institut f\"ur Genetik, Universit\"at zu K\"oln, 50674 K\"oln, Germany. (5)  CNAG-CRG, Centre for Genomic Regulation (CRG) and Universitat Pompeu Fabra (UPF), Barcelona, Spain. (6) Centre for Research in Agricultural
Genomics (CRAG), 08193 Bellaterra, Spain.

\normalsize

\begin{abstract}
We present an exact, closed expression for the expected neutral Site Frequency Spectrum for two neutral sites, 2-SFS,
without recombination. This spectrum is the immediate extension of the well known single site $\theta/f$ neutral
SFS.  Similar formulae are also provided for the case of the expected SFS of sites that are linked to
a focal neutral mutation of known frequency. Formulae for finite samples are obtained by coalescent methods and
remarkably simple expressions are derived for the SFS of a large population, which are also solutions of the
multi-allelic Kolmogorov equations. 
Besides the general interest of these new spectra, they relate to interesting biological cases such as structural
variants and introgressions. As an example, we present the expected neutral frequency spectrum of regions with
a chromosomal inversion.

\end{abstract}
\section{Introduction}

One of the major features that characterizes nucleotide polymorphisms is the Site Frequency Spectrum (SFS), that is the
distribution of the mutation frequencies at each site. The SFS can be computed either for the whole (large) population,
assuming that the frequency $f$ is a continuous value in $(0,1)$ or for a sample of $n$ individuals, for which the
frequency is a discrete variable  $f=k/n$, where $k\in[1,n-1]$. Typically sites with alleles at frequency 0 or 1 are not
included in the SFS. 

According to the standard neutral model of molecular evolution \citep{Kimura1983neutral}, polymorphisms segregating in a
population eventually reach a mutation-drift equilibrium. In this model, the expected neutral spectrum is proportional
to the inverse of the frequency \citep{wright1938distribution,ewens2012mathematical}.
Using coalescent theory, \cite{fu1995statistical} derived the mean and covariance matrix for each bin of the sample SFS, by
averaging coalescent tree realizations across the whole tree space. For a single realization of the coalescent tree,
results are different and depend on the realization; for example, mutations of high frequencies can be present only for highly
unbalanced genealogies \citep{ledda2015decomposing}. The SFS was also studied in scenarios including selection
\citep{Fay:2000tg,Kim:2002xe}, demography \citep{Griffiths:1994jt,vzivkovic2008second} or population structure
\citep{alcala2016genomic}.

Besides its general interest, the SFS has been used to devise goodness-of-fit statistical tests to estimate the
relevance of the standard neutral model for an observed dataset. SFS-based neutrality tests
contrast estimations of the nucleotide variability from different bins of the sample SFS
\citep{Tajima:1989dk,Fu:1993qm,Achaz:2009qp}. \cite{Ferretti:2010ss} showed that, once the SFS under
an alternative scenario (e.g. selection, demography or structure) is known, the optimal test to reject the standard
neutral model is based on the difference between the standard neutral SFS and the  alternative scenario SFS. All these
tests assume complete linkage among variants in their null model.

 Assuming independence between the sites, the observed SFS can also be used to estimate model parameters. An interesting
 recent approach is the estimation of piece-wise constant demography from genomewide SFS (e.g. \cite{liu2015exploring}).
 More sophisticated methods based on the expected SFS, such as Poisson Random Field
 \citep{sawyer1992population,Bustamante:2001df,Bustamante:2002dz} and Composite Likelihood approaches
 \citep[\textit{e.g., }][]{Kim:2002xe,Li:2005zv,Kim:2004sj,Nielsen:2005bs}, have also played an important role in the
 detection of events of selection across regions of the genome. However, the assumption of linkage equilibrium is often
 violated in genetic data. In fact, while the average spectrum is insensitive to recombination, the presence of linked 
 variants affects the distribution of summary statistics, therefore the spread (and possibly the mean) of the estimated
 parameters \citep{hudson1990gene,Thornton:2005il}. For this reason, simulations of the evolution of linked sequences 
 are required for an accurate estimation of the statistical support for different models \citep{Gutenkunst:2009ad}.

The joint SFS for multiple sites has been the subject of longstanding investigations. The simplest spectrum for multiple
sites is the ``two-locus frequency spectrum'' \citep{hudson2001two}, which we name the ``two-Sites Frequency Spectrum''
or 2-SFS. Assuming independence between the sites (\textit{i.e.} free recombination), it simply reduces to the random
association between two single-sites spectra (1-SFS). For intermediate recombination, a recursion solvable for small sample size has been provided \citep{Golding:1984oy,ethier1990two}  as a well as a numerical solution relying on simulations \citep{hudson2001two}. Without recombination, finding an analytical expression for the spectrum has proven to be difficult.

There is a close relation between the $m$-SFS (the joint SFS of $m$ sites) and the multi-allelic spectrum of a single
$locus$ (defined as a sequence with one or more sites). Under the \textit{infinite-sites} model, sites are assumed to
have at most two alleles as new mutations occur exclusively at non-polymorphic sites. At the locus scale, each haplotype
(the specific combination of the alleles carried at each point) can be interpreted as a single allele at a multi-allelic
locus.
In the absence of recombination, each point mutation either leaves the number of different haplotypes
unchanged or generates one new haplotype.
Therefore, at least conceptually, the SFS for $m$ biallelic sites at low mutation rate is closely related to the spectrum of $m+1$ alleles in a multi-allelic locus. Indeed, it is possible to retrieve the latter from the former by considering the $m+1$ alleles that result from the $m$ polymorphic sites. However, the  $m$-SFS contains extra-information on the different linkage between the sites that is not available in the multi-allelic locus spectrum.


For an infinite population, the multi-alleles single-locus spectrum is the solution of a multiallelic diffusion equation
\citep{ewens2012mathematical}. Polynomial expansions were proposed to solve the diffusion equations for the
SFS of an infinite population \citep{kimura1956random,littler1975transition,griffiths1979transition}. Finally, a polynomial expansion of the 2-SFS has been
found for two sites without recombination and with general selection coefficients \citep{xie2011site}. However, the
reported solution is an infinite series and is in sharp contrast with the simplicity of the solution for a single
neutral site: $\ev[\xi(f)] = \theta/f$. Furthermore, no closed form was provided for the 2-SFS of a sample.

Using a coalescent framework, the probability and size of two nested mutations were expressed by
\cite{hobolth2009genealogy} as sums of binomial coefficients. Their formulae can be rewritten as an expected SFS in terms of a finite series. However their conditioning on exactly two nested mutations skews the
spectrum and simulations show that even under this condition their result is valid only for 
$L\theta\ll 1$. Interesting analytical results on the spectrum of tri-allelic loci and recurrent mutations were obtained by Song and collaborators \citep{jenkins2011effect,jenkins2014general} for the Kingman coalescent and general allelic transition matrices. More recently, \cite{Sargsyan:2015hw} generalized the result of \cite{hobolth2009genealogy} by conditioning on any two mutations
(nested or not) and extending it to populations of variable size. Moreover, he clarified the notion and classification of the 2-SFS.

In this work, we present a simple closed-form solution for the expectation of the neutral 2-SFS without recombination, for both the discrete sample  2-SFS and the continuous population 2-SFS. 

The solution for a finite sample is obtained in a coalescent framework \citep{fu1995statistical,ferretti2012neutrality} and its extrapolation
to the limit of infinite sample sizes yields the continuous spectrum.
Furthermore, we derive the expected 1-SFS of sites that are completely linked to a focal mutation of known frequency. In the appendices we also extend our results on the 2-SFS into closed expressions for the multi-allelic spectrum of a locus with three alleles. 

Finally, as an application, we present exact results for the expected spectrum of neutral, non-recombining inversions.
Chromosomal inversions are structural variants that play an important role in the adaptive evolution of some species
\citep{hoffmann2004chromosomal}, the most well-known case being flies in the \emph{Drosophila} genus
\citep{krimbas1992drosophila,corbett2012population}. We derive the expected frequency spectrum of neutral mutations
linked to a neutrally evolving chromosomal inversion or a structural variant with similar properties. The neutral spectrum of inversions is more complex than the usual site frequency spectrum and represents the null model to detect population genetics signatures of selection on chromosomal variants \citep{kennington2006patterns,white2009population}.


\subsubsection*{Model definition and notation}

We consider a population of size $N$ of haploid individuals without recombination. All subsequent results can be applied
to diploids, provided that $2N$ is used instead of $N$, and to other cases by substituting the appropriate effective
population size. We denote by $\mu$ the  mutation rate per site  and by $\theta=2N\mu$ the population-scaled mutation
rate per site. We work in the infinite-sites approximation, that is valid in the limit of small mutation rate $\theta
\ll 1$. More properly, our results are derived in the limit $\theta\rightarrow 0$ with fixed non-zero $\theta L$, where
$L$ is the length of the sequence. The expected value $\ev[.]$ denotes the expectation with respect to the realizations
of the evolutionary process for the sequences in the sample or in the whole population. We use {\em mutation} as a synonym for derived allele.

\subsubsection*{Connection between sample and population SFS}

We denote by $\xi(f)$ the \emph{density} of mutations at frequency $f$ in the whole population and by $\xi_k$ the
\emph{number} of mutations at frequency $k/n$ in a sample of size $n$. Importantly, in both cases $f$ or $k$ refer to the frequency of the mutation, \textit{i.e.} of the \emph{derived} allele, and thus $\xi$ corresponds to the \emph{unfolded} SFS.

The two spectra (sample and population) are related. Assuming that a mutation has frequency $f$ in the population, the probability of having $k$ mutant alleles in a random sample of size $n$ is simply given by the Binomial $ {n \choose k} f^k(1-f)^{n-k}$. As the expected density of mutations at frequency $f$ in the population is given by $\ev[\xi(f)]$, one can easily derive the sample frequency from the population frequency using the following sampling formula:
\beq
\ev[\xi_k]=\int_{\frac1N}^{1-\frac 1N}  {n \choose k} f^k(1-f)^{n-k}~\ev[\xi(f)]\label{limit}~df
\eeq
assuming that $n\ll N$.

Conversely, the population SFS can be derived from the sample SFS using the limit of large sample size $n\rightarrow \infty$. For a sample of $n$ individuals, the interval between the frequency bins is $1/n$ and therefore the density of mutations at the continuous frequency  $f = k/n$ can be approximated\footnote{More formally, eq.(\ref{eqlim}) can be obtained from eq.(\ref{limit}) under the assumptions that $\frac1N\ll f,1-f$ and that the population SFS is smooth over a range of frequencies $\Delta f\sim \frac1N$. } by $\ev\left[\xi\left(\frac{k}{n}\right)\right]\approx \frac{\ev[\xi_k]}{1/n}=n\ev[\xi_k]$. The expected population spectrum can then be constructed from the limit:
\beq
\ev[\xi(f)]=\lim_{n\rightarrow\infty} n\ev[\xi_{\lfloor nf \rfloor}]\label{eqlim}
\eeq
for frequencies not too close to $\frac1N$ or $1-\frac1N$. 

For a sample of size $n$, the expected neutral spectrum for constant population size is $\ev[\xi_k]=\theta L/k$ and consequently, we have $\ev[\xi(f)]=\theta L/f$ \citep{wright1938distribution,ewens2012mathematical}. These results are exact for the Kingman coalescent and the diffusion equations respectively, and they are approximately valid for neutral models for frequencies $f\gg\frac1N$. For frequencies of order $\frac1N$, model-dependent corrections are needed and equation (\ref{eqlim}) is not valid anymore.


In the rest of this section we will deal with sample and population spectra together. We will slightly abuse the notation and switch between number and density of mutations, or probability and probability density.

\subsubsection*{Conditional 1-SFS and joint 2-SFS}

In the following, we will use two related but different kinds of spectra. 

The first one is the joint 2-SFS of two bi-allelic sites. It is denoted $\xi(f_1,f_2)$ for the population and $\xi_{k,l}$ for the sample. It is defined as the density of pairs of sites  with mutation frequencies at $f_1$ and $f_2$ for the population (resp. $k/n$ and $l/n$ for the sample). This is a natural generalization of the classical SFS for a single site. The expected spectrum $\ev[\xi(f_1,f_2)]$ has two equivalent interpretations in the small $\theta$ limit: (a) for a sequence, it is the expected density of pairs of sites that harbor mutations with frequencies $f_1$ and $f_2$; (b) for two randomly chosen linked polymorphic sites, it is the probability density that they contain mutations with frequencies $f_1$ and $f_2$. 

The second one is a conditional 1-SFS, a frequency spectrum of sites that are linked to a focal mutation of frequency
$f_0$.
It is denoted $\xi(f|f_0)$ for the population and $\xi_{k|l}$ for the sample.  Again, this spectrum represents both (a) 
the expected density of single-site mutations of frequency $f$ in a locus linked to a focal neutral mutation of frequency $f_0$ and (b) the probability density that a randomly chosen site (linked to the focal site) hosts a mutation at frequency $f$.

Note that despite the similarity in notation, the two spectra $\xi(f,f_0)$ and $\xi(f|f_0)$ are different. The difference is the same as the one between the \emph{joint probability} $p(f,f_0)$ that two sites $x$ and $x_0$ have mutations of frequency $f$ and $f_0$ respectively, and the \emph{conditional probability} $p(f|f_0)$ that a mutation at site $x$ has frequency $f$ given that there is a mutation of frequency $f_0$ at a focal linked site $x_0$. Furthermore, the joint spectrum $\xi(f,f_0)$ refers to pairs of sites -- \textit{i.e.} it is a 2-SFS -- while the spectrum of linked sites $\xi(f|f_0)$ is a single-site SFS.

The relation between both types of spectra can be understood from the relation between the probabilities. The expected spectrum $\ev[\xi(f)]$ is given by the probability to find a mutation of frequency $f$ at a specific site, multiplied by the length of the sequence: $\ev[\xi(f)]=p(f)L$. As noted above, when $L=1$ (i.e. a locus with a single site is considered), $\ev[\xi(f)]$ corresponds to a proper probability $p(f)$. Assuming the presence of a mutation of frequency $f_0$ at a focal site, we have $\ev[\xi(f|f_0)]=p(f|f_0)(L-1)$. For pairs of sites, the expected number of mutations at frequencies $(f,f_0)$ is $\ev[\xi(f,f_0)]=p(f,f_0)L(L-1)$ when $f\neq f_0$ or $p(f_0,f_0)L(L-1)/2$ when $f=f_0$. The additional factor $\frac{1}{2}$ accounts for the symmetrical case of equal frequencies $f=f_0$. The equality $p(f,f_0)=p(f|f_0)p(f_0)$ applied to sample and population spectra, results in the following relations:
\beq
\ev[\xi_{k,l}]=\frac{\ev[\xi_{k|l}]\cdot \ev[\xi_l]}{1+\delta_{k,l}}=\begin{cases}\ev[\xi_{k|l}]\cdot \ev[\xi_l] & \mathrm{for}\ k\neq l\\ \frac{1}{2}\cdot\ev[\xi_{k|l}]\cdot \ev[\xi_l] & \mathrm{for}\ k=l \end{cases}\label{eq_spec_prob_s}
\eeq
\beq
\ev[\xi(f,f_0)]=\frac{\ev[\xi(f|f)]\cdot \ev[\xi(f)]}{1+\delta_{f,f_0}} =\begin{cases}\ev[\xi(f|f_0)]\cdot \ev[\xi(f_0)] & \mathrm{for}\ f\neq f_0\\ 
\frac{1}{2} \cdot \ev[\xi(f|f)]\cdot \ev[\xi(f)] & \mathrm{for}\ f= f_0\end{cases}\label{eq_spec_prob_p}
\eeq
where 
 $\delta_{x,y}$ is 1 if $x=y$, and 0 otherwise. Note that $x$ and $y$ can be either discrete or continuous variables. 

By definition, the 2-SFS includes only pairs of sites that are \emph{both} polymorphic. The probability that a pair of
sites contains a single polymorphism of frequency $k/n$ depends only on the 1-SFS and it is approximately equal to $2\ev[\xi_k]$ for $\theta\ll 1$. Consequently, on a sequence of size $L$ hosting $S$ polymorphic sites, the number of pairs of sites for which only one of the two is polymorphic of frequency $k/n$ is $\ev[(L-S)\xi_k]=L\cdot\ev[\xi_k]-\ev[S\xi_k]\approx L\cdot\ev[\xi_k]$ for small $\theta$.

\section{Results}


\subsection{Decomposition of the 2-SFS}
We follow \cite{Sargsyan:2015hw} and divide the 2-SFS $\xi(f_1,f_2)$ without recombination into two different components:
one \emph{nested} component $\xi^{N}(f_1,f_2)$ for cases where there are individuals carrying the two mutations (one is ``nested'' in the other), and a \emph{disjoint} component $\xi^{D}(f_1,f_2)$ that includes disjoint mutations only present in different individuals. The overall spectrum is  given by:
\begin{eqnarray}
	\xi(f_1,f_2) & = & \xi^{N}(f_1,f_2)+\xi^{D}(f_1,f_2) \label{full_2sfsp}\\
	\xi_{k,l}        & = & \xi^{N} _{k,l}+\xi^{D}_{k,l} \label{full_2sfss}
\end{eqnarray}

It is noteworthy to mention that that the overall spectrum cannot fully describe the genetic state of the two sites, while the two components
$\xi^{N}(f_1,f_2)$, $\xi^{D}(f_1,f_2)$ give a complete description up to permutations of all the haplotypes, similarly
to the usual SFS for one site. For example, the following haplotypes (derived alleles marked in bold)
\begin{center}
	
\textit{C}\textit{T} $\quad$\phantom{and}$\quad$\textit{C}\textit{\textbf{A}} 

\textit{C}\textit{\textbf{A}} $\quad$and$\quad$\textit{C}\textit{\textbf{A}}

\textit{\textbf{G}}\textit{\textbf{A}} $\quad$\phantom{and}$\quad$\textit{\textbf{G}}\textit{T} 
\end{center}
are identical from the point of view of the overall two-loci spectrum: in both samples there is just a pair of mutations
with allele count 1 and 2 respectively, therefore the only (symmetrical) nonzero value of the spectrum is
$\xi_{1,2}=\xi_{2,1}=1$. However the samples can be distinguished by the two components, since in the first one
the mutations are nested ($\xi^N_{1,2}=\xi^N_{2,1}=1$), while in the second one they are disjoint
($\xi^D_{1,2}=\xi^D_{2,1}=1$).
For this reason, these two components constitute the core of the two-loci SFS.

\begin{figure}[h]
\begin{center}
\includegraphics[width=\textwidth]{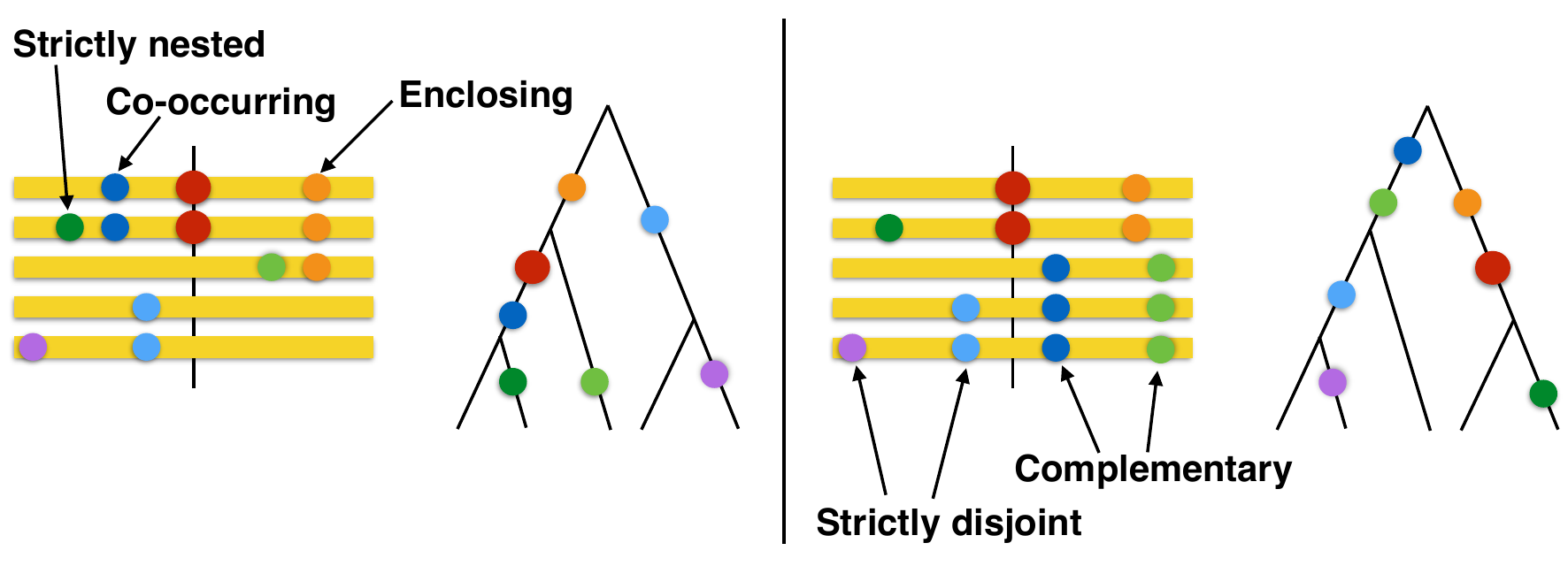}
\caption{An illustration of two non-recombining loci and their corresponding genealogical trees. The yellow segments represent the ancestral sequence and the colored bullets represent derived alleles. This figure illustrates the classification of all possible types of mutations with respect to the focal mutation (in red) and their occurrence on the sequence tree. Nested mutations are indicated in the left panel, disjoint mutations in the right one.\\
If the focal mutation is not on a root branch (left), it is clear from the figures that mutations can be on the same branch as the focal mutation (\emph{co-occurring}), on the subtree below (\emph{strictly nested}), between the focal mutation and the root (\emph{enclosing}), or on other branches (\emph{strictly disjoint}). If the mutation is on a root branch (right), there cannot be enclosing mutations, but there are mutations on the other root branch (\emph{complementary}).}\label{fig_mut}
\end{center}
\end{figure}

Without recombination, the conditional 1-SFS  $\xi(f|f_0)$ can be also decomposed further\footnote{We subdivide the ``strictly nested'' mutations of \cite{Sargsyan:2015hw}
into \emph{strictly nested} and \emph{enclosing} mutations while we refer to his ``identical'' mutations as \emph{co-occurring}.} into different
subspectra. They are
illustrated in Figure \ref{fig_mut}:
\begin{itemize}
\item $\xi^{(sn)}(f|f_0)$ : \emph{strictly nested} mutations, where the mutation is carried only by a subset of individuals with the focal mutation;
\item $\xi^{(co)}(f|f_0)$ : \emph{co-occurring} mutations, where both mutations are systematically carried by the same individuals;
\item $\xi^{(en)}(f|f_0)$ : \emph{enclosing} mutations, where only  a subset of individuals with the mutation also carry the focal one;
\item $\xi^{(cm)}(f|f_0)$ : \emph{complementary} mutations, where each individual has only one of the two mutations;
\item $\xi^{(sd)}(f|f_0)$ : \emph{strictly disjoint} mutations, where the mutation is carried by a subset of the individuals without the focal one.
\end{itemize}

Importantly, without recombination, enclosing and complementary mutations cannot be present together in the same sequence. 

Given the rules of conditional probabilities $p(f,f_0)=p(f|f_0)p(f_0)$ and the interpretations above, the relations between the two sets of population subspectra are:
\begin{align}
\ev[\xi^{N}(f,f_0)]=&
\Big(\ev[\xi^{(sn)}(f|f_0)]+\ev[\xi^{(co)}(f|f_0)]+\ev[\xi^{(en)}(f|f_0)]\Big)\cdot \frac{ \ev[\xi(f_0)] }{
1+\delta_{f,f_0} }\label{eq1p} \\
\ev[\xi^{D}(f,f_0)]=&
\Big(\ev[\xi^{(cm)}(f|f_0)]+\ev[\xi^{(sd)}(f|f_0)]\Big)\cdot\frac{ \ev[\xi(f_0)]}{1+\delta_{f,f_0}} \label{eq2p}
\end{align}
Similarly, for sample spectra, we have
\begin{align}
\ev[\xi^{N}_{k,l}]=&
\left(\ev[\xi^{(sn)}_{k|l}]+\ev[\xi^{(co)}_{k|l}]+\ev[\xi^{(en)}_{k|l}]\right)\cdot\frac{ \ev[\xi_l] }{1+\delta_{k,l}}\label{eq1s} \\
\ev[\xi^{D}_{k,l}]=&
\left(\ev[\xi^{(cm)}_{k|l}]+\ev[\xi^{(sd)}_{k|l}]\right) \cdot\frac{ \ev[\xi_l] }{1+\delta_{k,l}} \label{eq2s}
\end{align}

\subsection{The joint and conditional SFS}

In this section, we report the conditional and joint spectra both for the sample and the population. The derivations and
proofs of all equations in this section are given in the Methods and Supplementary Material, as well as comparisons of the
analytical spectrum with simulations. The folded version of the 2-SFS  is provided
in Appendix \ref{app_folded} for completeness.

\subsubsection{The sample joint 2-SFS}

%

Using equations \ref{eq1s} and  \ref{eq2s}, one can derive the two components of the 2-loci spectrum as\footnote{Note
that the related formula (14) in the paper by \cite{ferretti2012neutrality} has a sign error. It should be identical to the second equation in (\ref{sfs_2s}) up to a multiplicative factor.}:
\begin{align}
\ev[\xi^{N}_{k,l}]=&\begin{cases}\theta^2 L^2 \frac{\beta_n(k)-\beta_{n}(k+1)}{2}\quad&
\mathrm{for}\ k<l\\
\theta^2 L^2 \frac{\beta_{n}(k)}{2} \quad& \mathrm{for}\ k=l\\
\theta^2 L^2 \frac{\beta_n(l)-\beta_{n}(l+1)}{2}\quad & \mathrm{for}\ k>l
\end{cases}
\nonumber
\\
\ev[\xi^{D}_{k,l}]=&\begin{cases}
\theta^2L^2\left(\frac{1}{kl}-\frac{\beta_n(k)-\beta_n(k+1)+\beta_n(l)-\beta_n(l+1)}{2}\right)\frac{2-\delta_{k,l}}{2}\quad & \mathrm{for}\ k+l<n\\
               \theta^2L^2\left(\frac{a_n-a_k}{n-k}+\frac{a_n-a_l}{n-l}-\frac{\beta_n(k)+\beta_n(l)}{2}\right) \frac{2-\delta_{k,l}}{2}\quad & \mathrm{for}\ k+l=n\\
 0 \quad & \mathrm{for}\ k+l>n
               \end{cases}\label{sfs_2s}
\end{align}
with
\beq
a_n=\sum_{i=1}^{n-1}\frac{1}{i}\quad,\quad \beta_n(i)=\frac{2n}{(n-i+1)(n-i)}(a_{n+1}-a_i)-\frac{2}{n-i} \nonumber 
\eeq

As shown by equation (\ref{full_2sfss}), the full spectrum is simply the sum of the two above equations.


\subsubsection{The population joint 2-SFS}

Similarly, the 2-SFS for the whole population is given by the sum of the two following equations:
\begin{align}
\ev[\xi^{N}(f,f_0)]=  & \theta^2 L^2 \cdot\left[\frac{1}{(1-\min(f,f_0))^2}\left(1+\frac{1}{\min(f,f_0)}+\frac{2\ln(\min(f,f_0))}{1-\min(f,f_0)}\right) \right.\nonumber  \\
                              + & \left.\delta(f-f_0)\frac{f_0}{1-f_0}\left(-\frac{\ln(f_0)}{1-f_0}-1\right)\right] \nonumber\\
\ev[\xi^{D}(f,f_0)]= & \theta^2 L^2  \cdot\left[ \frac{1}{ff_0}-\frac{1}{(1-f)^2}\left(1+\frac{1}{f}+\frac{2\ln(f)}{1-f}\right) -\frac{1}{(1-f_0)^2}\left(1+\frac{1}{f_0}+\frac{2\ln(f_0)}{1-f_0}\right)\right.\nonumber\\
+ & 
\left. \delta(f-1+f_0)\left(\frac{1-f_0}{f_0^2}\ln(1-f_0)+\frac{f_0}{(1-f_0)^2}\ln(f_0) +\frac{1}{f_0(1-f_0)}\right)\right]\label{sfs_2p}
\end{align}

with $\ev[\xi^{N}(f,f_0)]=0$ for $f>f_0$ and $\ev[\xi^{D}(f,f_0)]=0$ for $f+f_0>1$. 

Here, we denote by $\delta(f-f_0)$ the density of the Dirac ``delta function'' distribution concentrated in $f_0$ (i.e. $\delta(f-f_0)=0$ for $f\neq f_0$, normalized such as $\int^\infty_{-\infty}\delta(f-f_0) df=1$).

\subsubsection{The sample conditional 1-SFS}
The conditional 1-SFS for sites that are linked to a focal mutation of count $l$ is simply the sum of all its components, given by the following equations:
\begin{align}
\ev[\xi^{(sn)}_{k|l}]=&\theta L\cdot l\frac{\beta_n(k)-\beta_{n}(k+1)}{2}\quad\mathrm{for}\ k<l \nonumber\\
\ev[\xi^{(co)}_{k|l}]=&\theta L \cdot l\beta_n(k) \delta_{kl} \nonumber\\
\ev[\xi^{(en)}_{k|l}]=&\theta L \cdot l\frac{\beta_n(l)-\beta_{n}(l+1)}{2} \quad\mathrm{for}\  k>l\label{sfs_ls}\\
\ev[\xi^{(cm)}_{k|l}]=&\theta L \cdot l \left(\frac{a_n-a_k}{n-k}+\frac{a_n-a_l}{n-l}-\frac{\beta_n(k)+\beta_n(l)}{2}\right)\delta_{k,n-l}\nonumber\\
\ev[\xi^{(sd)}_{k|l}]=&\theta L \cdot \left( \frac{1}{k}-l\frac{\beta_n(k)-\beta_n(k+1)+\beta_n(l)-\beta_n(l+1)}{2}\right)\quad\mathrm{for}\  k+l<n\nonumber
\end{align}
and 0 otherwise.


\subsubsection{The population conditional 1-SFS}

For the whole population, this becomes:
\begin{align}
\ev[\xi^{(sn)}(f|f_0)]= & \theta L \cdot\frac{f_0}{(1-f)^2}\left(1+\frac{1}{f}+\frac{2\ln(f)}{1-f}\right) 
,\quad f<f_0 \nonumber\\
\ev[\xi^{(co)}(f|f_0)]= &
\theta L  \cdot\delta(f-f_0)\frac{2f_0}{1-f_0}\left(-\frac{\ln(f_0)}{1-f_0}-1\right) \nonumber\\
\ev[\xi^{(en)}(f|f_0)]= & \theta L  \cdot\frac{f_0}{(1-f_0)^2}\left(1+\frac{1}{f_0}+\frac{2\ln(f_0)}{1-f_0}\right) 
,\quad f>f_0 \label{sfs_lp}\\
\ev[\xi^{(cm)}(f|f_0)]= & 
\theta L \cdot \delta(f-1+f_0)\left[\frac{1-f_0}{f_0}\ln(1-f_0)+\left(\frac{f_0}{1-f_0}\right)^2\ln(f_0) +\frac{1}{1-f_0}\right] \nonumber\\
\ev[\xi^{(sd)}(f|f_0)]= & \theta L  \cdot\left[ \frac{1}{f}-\frac{f_0}{(1-f)^2}\left(1+\frac{1}{f}+\frac{2\ln(f)}{1-f}\right)\right. \nonumber \\ & \left. -\frac{f_0}{(1-f_0)^2}\left(1+\frac{1}{f_0}+\frac{2\ln(f_0)}{1-f_0}\right)\right]
,\quad f<1-f_0\nonumber
\end{align}

\subsection{Shape of the SFS}

\begin{figure}[h]
\begin{center}
\includegraphics[width=14cm]{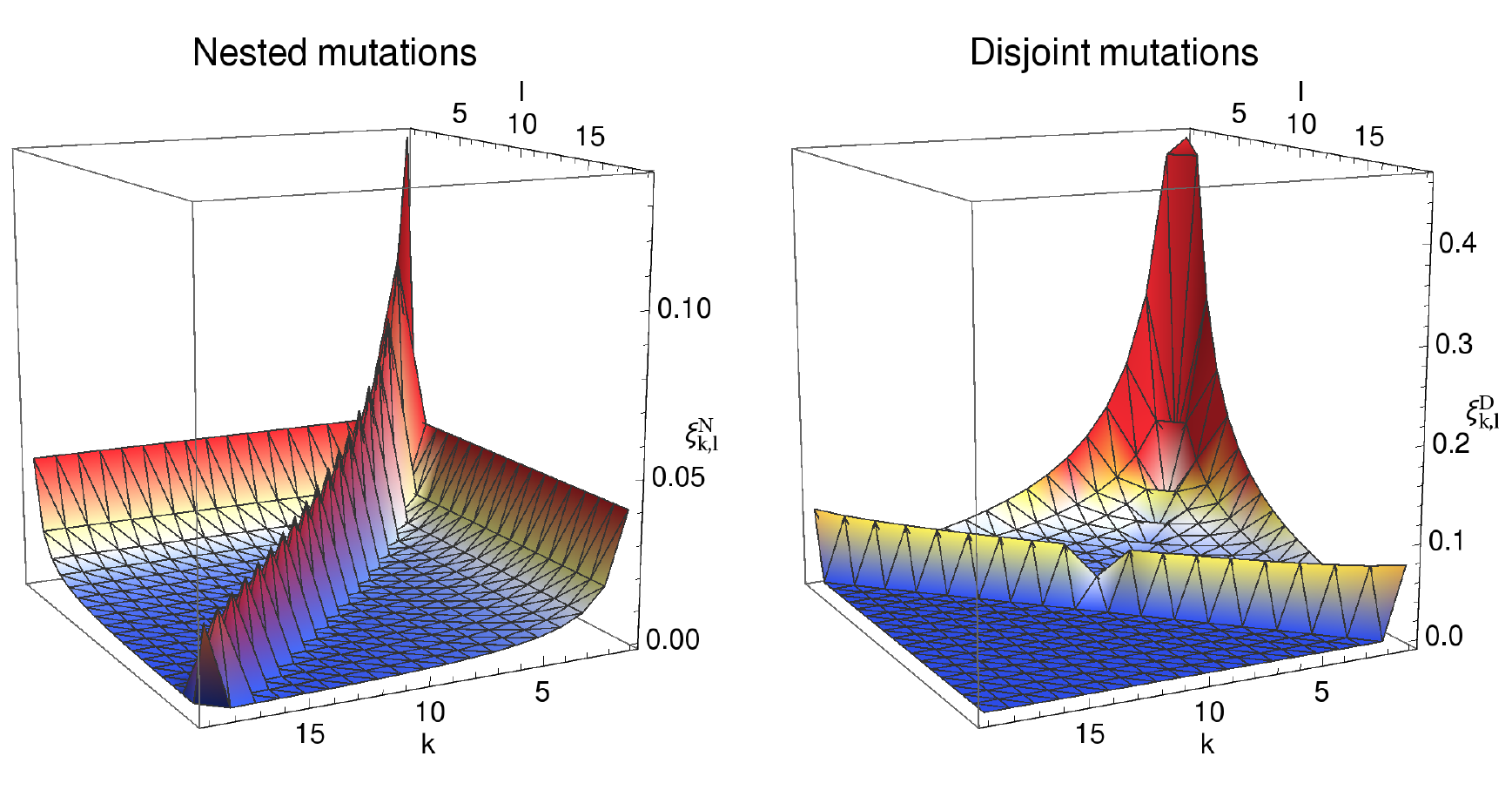}
\caption{Plots of nested and disjoint contributions to the two-locus frequency spectrum for $\theta L=1$, $n=20$. Note the different scales of the two plots.}\label{fig_sfs2}
\end{center}
\end{figure}

We report the full joint 2-SFS as well as both the nested and disjoint component (Figure  \ref{fig_sfs2}). Nested mutations have preferentially a rare mutation in either site -- so that the mutation at lower frequency is easily nested into the other -- or are co-occurring mutations -- corresponding to mutation found in the same branch of the genealogical tree. Disjoint mutations are dominated by cases where both mutations are rare -- mostly disjoint -- or by complementary mutations.  The large contribution of co-occurring (nested component) and complementary mutations (disjoint component) is a direct consequence of the two long branches that coalesce at the root node of a Kingman tree.

\begin{figure}[h]
\begin{center}
\includegraphics[width=4.4cm]{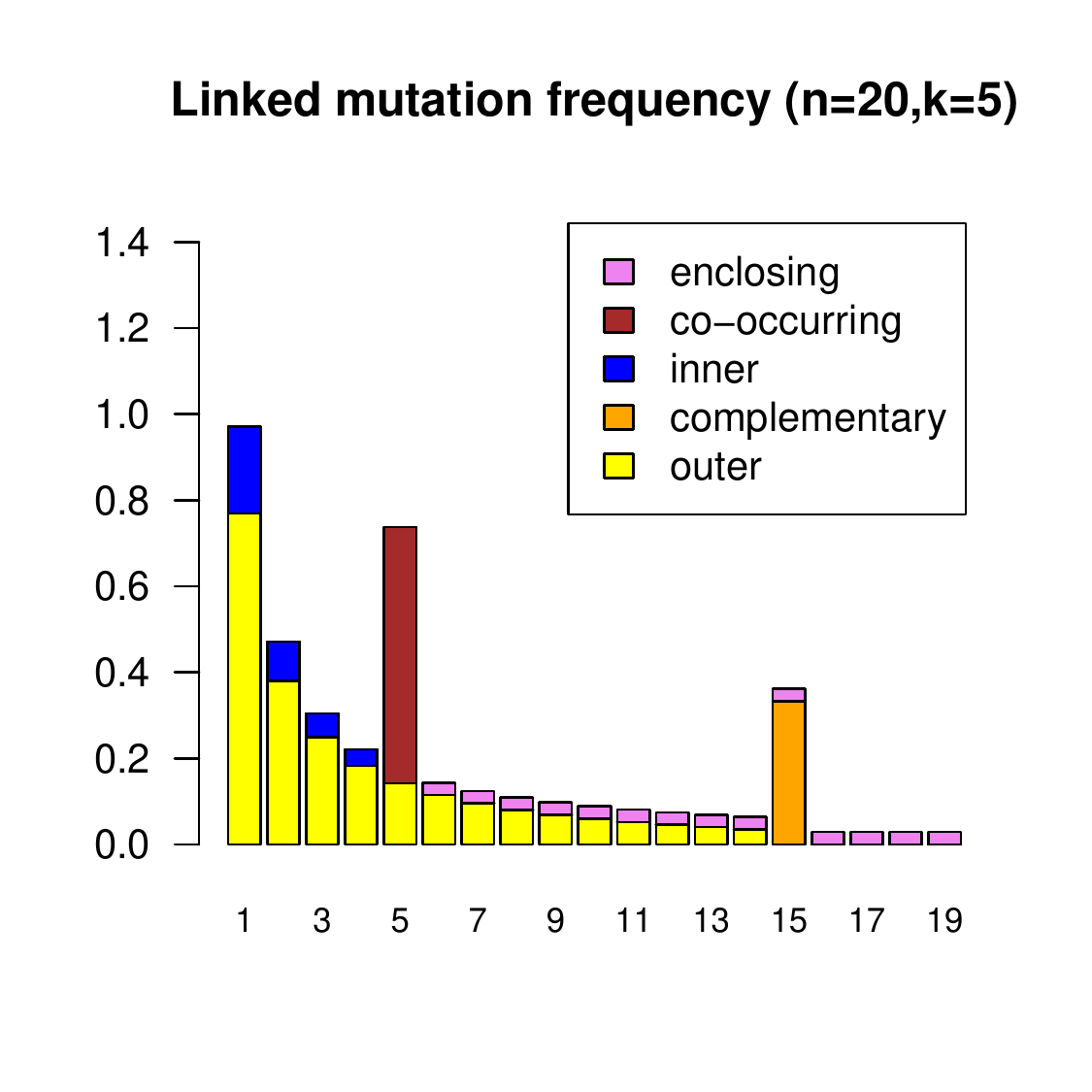}
\includegraphics[width=4.4cm]{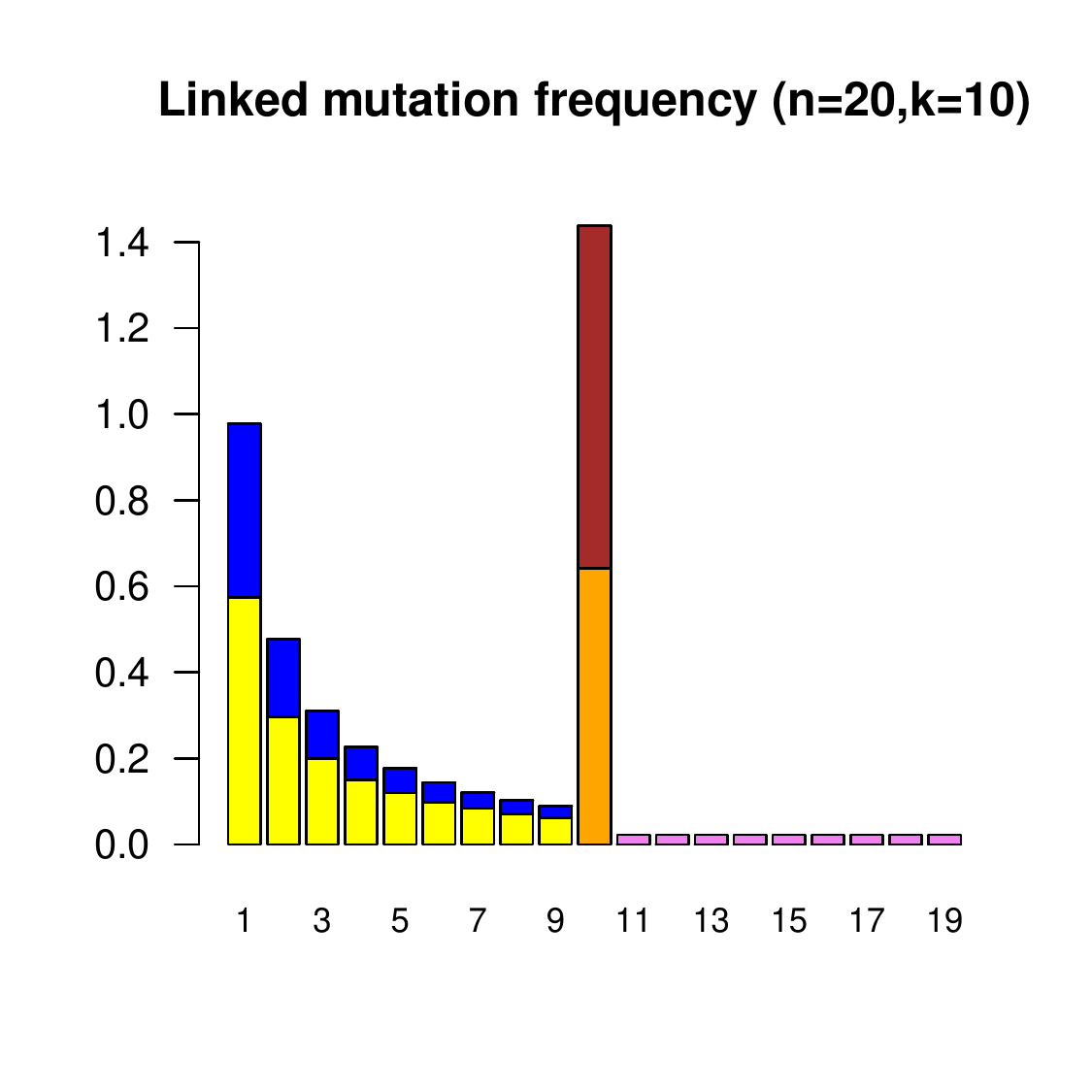}
\includegraphics[width=4.4cm]{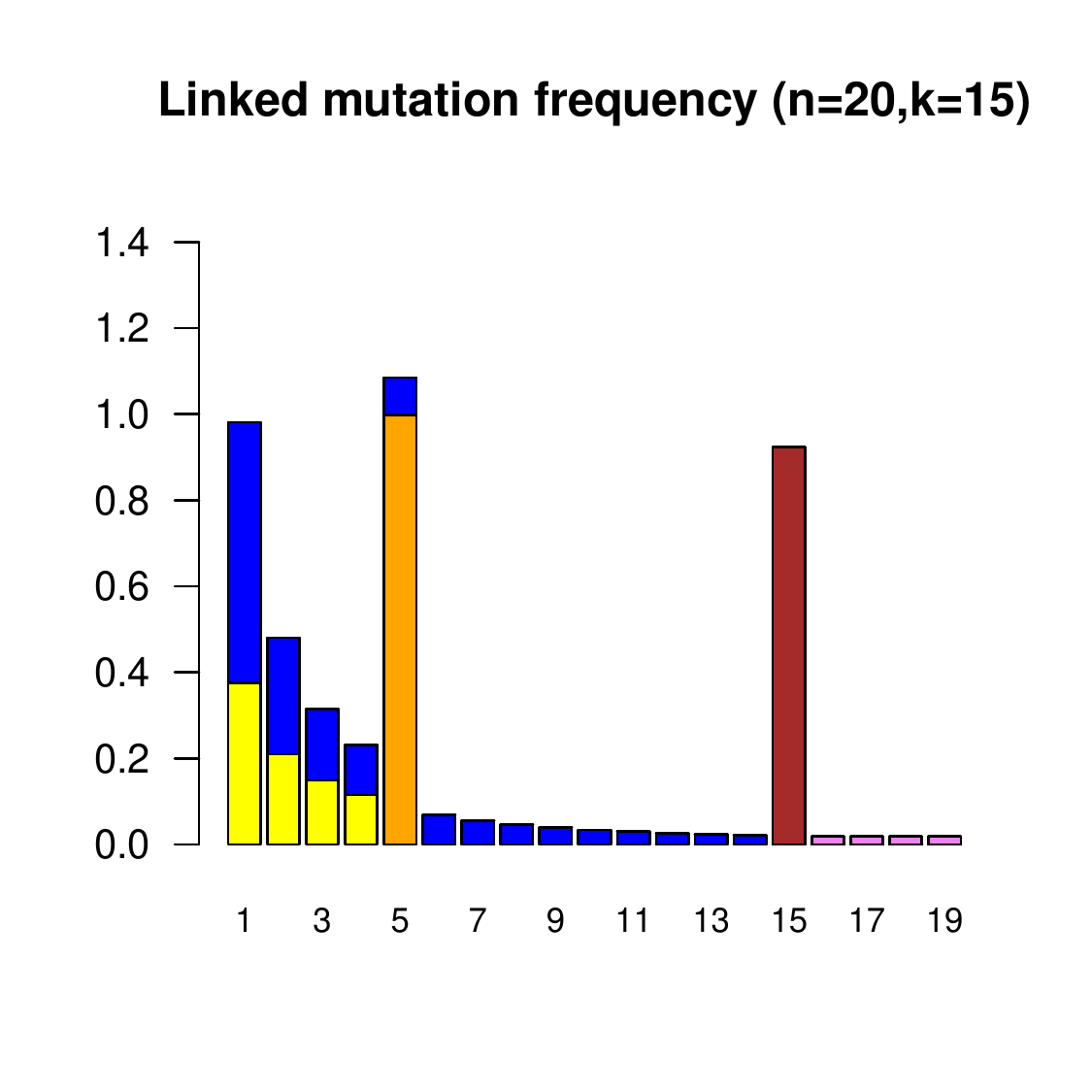}
\caption{  Barplot of the spectrum of linked sites for $\theta L=1$, $n=20$, each column colored according to the
different contributions. The focal mutation has frequency 5/20=0.25 (left), 10/20=0.5 (middle) and 15/20=0.75 (right)
respectively.}\label{fig_sfsl}
\end{center}
\end{figure}

The conditional 1-SFS of linked sites and the relative contributions of each component to each frequency are shown in Figure \ref{fig_sfsl}. Co-occurring and complementary mutations also account for a considerable fraction of the spectrum, especially when the focal mutation ($f_0$) is at high frequency. The rest of the spectrum is biased towards mutations with a lower frequency than the focal one. Strictly nested mutations are important only when the frequency of the focal mutation is intermediate or high. Enclosing mutations are typically negligible and their abundance is uniform as it was also noticed by \cite{hobolth2009genealogy}.

\begin{figure}[h]
\begin{center}
\includegraphics[width=10cm]{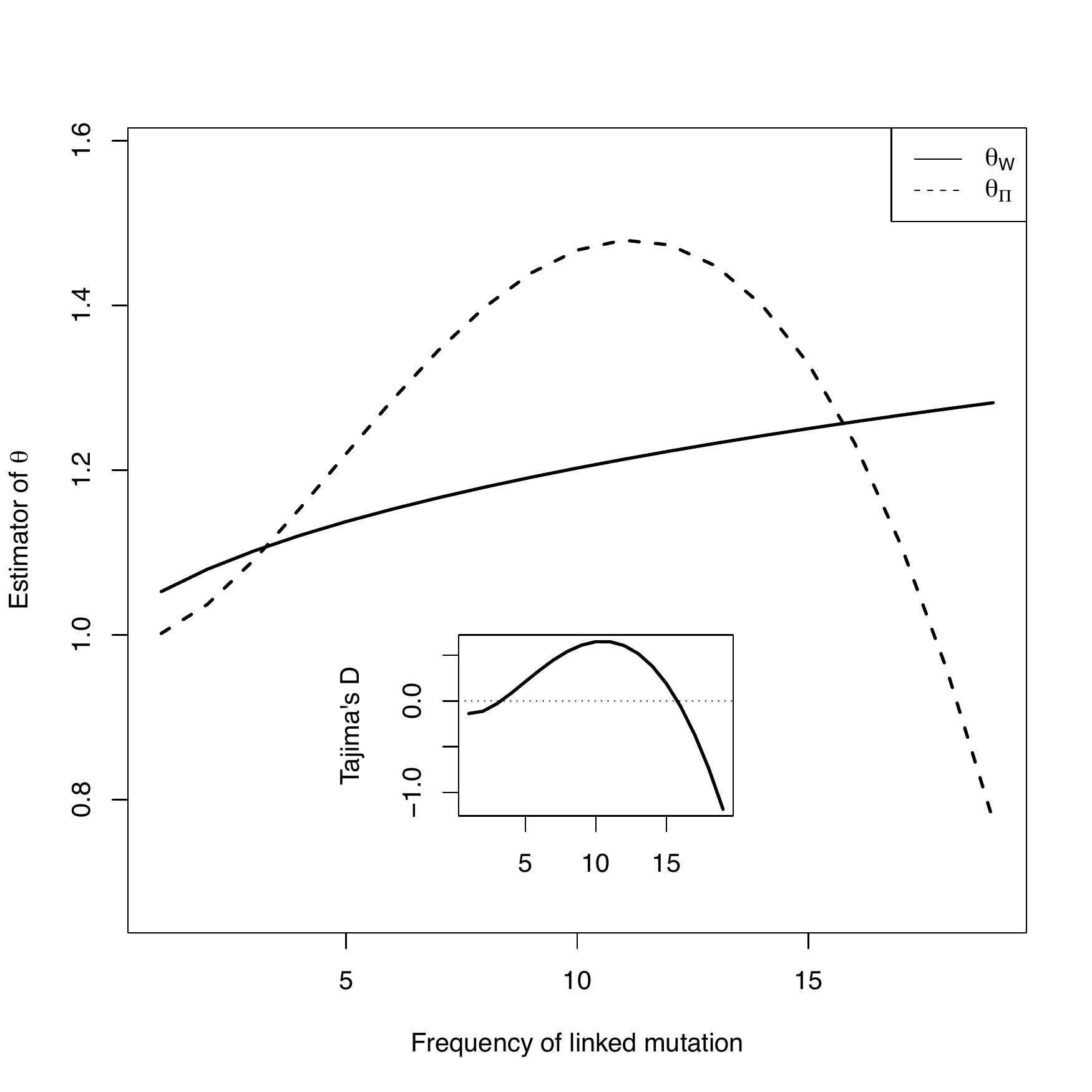}
\caption{Mean values of the Watterson estimator ($\hat \theta_S$) and Tajima estimator ($\hat \theta_\pi$) of $\theta$ conditioned on the presence of a linked mutation, for $\theta=1$, $n=20$. In the inset, approximate mean value of Tajima's $D$ (computed substituting $S$ with its mean value in the denominator).}\label{fig_wattpi}
\end{center}
\end{figure}

Finally, in Figure \ref{fig_wattpi} we show the impact of having a focal mutation of a known frequency on two estimators
of $\theta$. The \cite{watterson1975number} estimator, $\hat \theta_S$, depends on the total number of polymorphic
sites, which increases with the frequency of the focal mutation, while \cite{tajima1983evolutionary} estimator, $\hat
\theta_\pi$, is more sensitive to mutations of intermediate frequency. Therefore the comparison between the two
illustrates how the spectrum is skewed towards common or rare mutations. As Tajima's $D$ \citep{Tajima:1989dk} is proportional to the difference $\hat \theta_\pi - \hat \theta_S$, positive values for this test statistic suggest an excess of common mutations while negative values point to an excess of rare mutations. Figure \ref{fig_wattpi} shows that the spectrum has a slight excess of rare mutations at low frequencies of the focal mutation and an excess of common mutations for intermediate frequencies, while it is dominated again by rare mutations if the focal mutation is at high frequencies.


\subsection{The frequency spectrum of chromosomal inversions}
Chromosomal inversions are chromosomal rearrangements in which the orientation of a segment of a chromosome gets
reversed. They are well known structural variants, sometimes with important phenotypic effects. Recombination between
normal and inverted sequences is strongly suppressed due to mechanical incompatibilities during crossing over, selection
against unbalanced chromosomes and presumably other, still unknown, reasons \citep{Kirkpatrick2010}.

An inversion does not destroy the genetic information of the sequence, but adds a new ``allelic'' component given
by the orientation of the sequence. Apart from the inhibition of recombination, this orientation ``allele'' is for
our purposes akin to a normal point mutation of the same frequency. This is true also for its evolution. Hence, the expected spectrum of neutral inversions can be derived from our results for the linked spectrum, considering the orientation of the sequence as the focal mutation. 

\begin{figure}[h]
\begin{center}
\includegraphics[width=\textwidth]{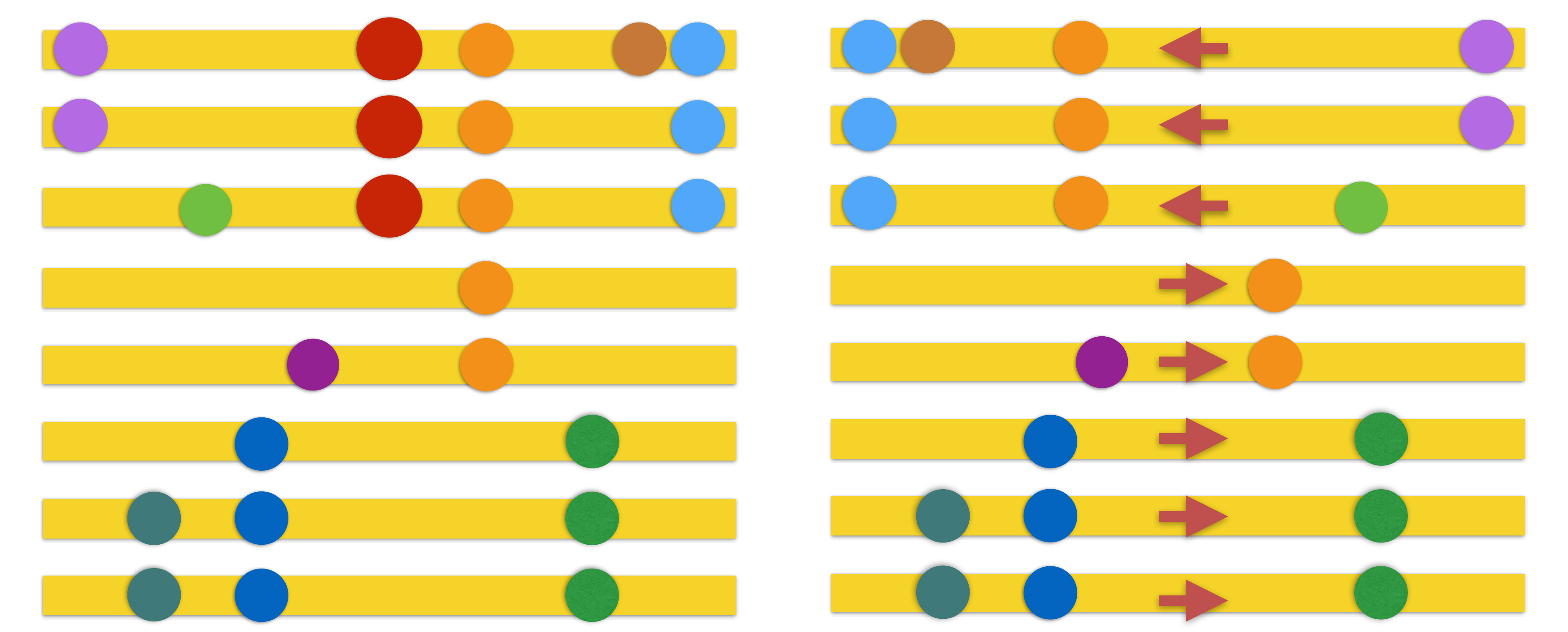}
\caption{Illustration of the similarity between inversions (right) and SNPs (left). The yellow segments represent the ancestral sequence, the red arrows represent its orientation, while the colored bullets represent derived SNPs.}\label{fig_inv}
\end{center}
\end{figure}

If we assume that the original orientation of the
sequence is known (e.g. by synteny with a close species) and that the orientation of the sequence is known for all individuals in the sample, then the spectrum of inversions has the same components as the spectrum of sites linked to a focal
mutation, as illustrated in Figure \ref{fig_inv}. This is a consequence of the suppression of recombination between normal and inverted alleles. 

We denote the sample spectrum of inversions by $\mathcal{I}_{k|i}$ where $k/n$ is the frequency of mutations and $i/n$ is the frequency of the inversion. If we assume that the rate of inversions is low, i.e. that multiple segregating overlapping inversions are unlikely to occur, then the inversion follows the infinite-sites model. Moreover, recombination within normal or inverted sequences does not affect the joint spectrum of the inversion and a point mutation therein, because it does change not their frequency, nor their linkage. Hence, the expected spectrum of neutral inversions follows directly from our results on the conditional 1-SFS:
\beq
\mathcal{I}_{k|i}^{(sn)}=\xi_{k|i}^{(sn)}\ ,\ \mathcal{I}_{k|i}^{(co)}=\xi_{k|i}^{(co)}\ ,\ \mathcal{I}_{k|i}^{(en)}=\xi_{k|i}^{(en)}\ ,\ \mathcal{I}_{k|i}^{(cm)}=\xi_{k|i}^{(cm)}\ ,\ \mathcal{I}_{k|i}^{(sd)}=\xi_{k|i}^{(sd)}
\eeq
The same applies to the population spectrum.


If the original orientation of the sequence is unknown, it could be inferred from the frequency spectrum by a
Bayesian approach similar to the one employed in \cite{Sargsyan:2015hw} for non-inverted haplotypes.

\section{Methods}



\subsection{The sample joint 2-SFS}

To obtain the sample spectrum for pairs of mutations, we notice that this spectrum can be defined in terms of the expected value of crossproducts of the usual SFS. In detail, we have
\beq
\ev[\xi_{k,l}]=\mathrm{E}[\xi_k\xi_l], \text{ if } k\neq l
\eeq
and
\beq
\ev[\xi_{k,k}]=\ev[\xi_k(\xi_k-1)]/2.
\eeq 

These expected values have been derived by \cite{fu1995statistical} by coalescent methods. However his results do not
distinguish the different contributions from nested and disjoint mutations to the spectrum.


Tracking the origin of each term in the derivation, it is easy to show that equations (24) and (28) of \cite{fu1995statistical} contribute to nested pairs of mutations, while equations (25), (29) and (30) contribute to disjoint pairs of mutations. All these terms combine linearly and do not interfere, therefore we can decompose the resulting $\mathrm{E}[\xi_k\xi_l]$ into contributions coming from equations (24),(28) and (25),(29) and (30) of \cite{fu1995statistical}.
This can be obtain directly by Fu's expression for the covariance matrix $\sigma_{kl}$, since $\mathrm{E}[\xi_k\xi_l]=\delta_{k,l} \mathrm{E}[\xi_k]+\mathrm{E}[\xi_k]\mathrm{E}[\xi_l]+\theta^2L^2\sigma_{kl}$ and $\mathrm{E}[\xi_k]=\theta L/k$. 

A detailed review of the calculations of \cite{fu1995statistical}, tracking the parts that lead to our mutation
classes, is provided in the Supplementary Material. 

The same results could also be  obtained by re-interpreting the results of \citet{jenkins2011effect} from Theorem 5.1 for small $\theta L$ ($\theta$ in their article). Their results for recurrent mutations are mathematically equivalent to the results for mutations in an infinite-sites model, for a special choice of  allele transition matrices (in the triallelic case, a strictly lower triangular matrix with all non-zero entries equal to 1). Their classification is based on the location of the mutations on the tree: their ``nested mutations'' correspond to strictly nested and enclosing mutations here, ``mutations on the same branch'' correspond to co-occurring mutations, ``mutations on basal branches'' correspond to complementary mutations, and ``non-nested mutations' correspond to strictly disjoint mutations.



\subsection{The sample conditional 1-SFS}

The spectrum for sites linked to a focal mutation of count $l$ (equation \ref{sfs_ls}) can be obtained from the previous spectrum (\ref{sfs_2s}). The first step is simply to condition on the frequency $l/n$ of the focal mutation, i.e. dividing the 2-SFS $\mathrm{E}[\xi_{k,l}]$ by $\mathrm{E}[\xi_l]\frac{1+\delta_{k,l}}{2}$ following equations (\ref{eq1s}) and (\ref{eq2s}). In fact, $\ev[\xi_{k|l}]=(L-1)\mathrm{P}[c(x)=k|c(y)=l]=L(L-1)\mathrm{P}[c(x)=k,c(y)=l]/L\mathrm{P}[c(y)=l]=\frac{2}{1+\delta_{k,l}}\ev[\xi_{k,l}]/\ev[\xi_l]$ where $c(x)$ is the derived allele count at site $x$. 

The second step is to break further the two contributions of the resulting conditional spectrum into the different components. Strictly nested, co-occurring and enclosing mutations are derived from the nested contribution and are distinguished by site frequencies only:  strictly nested ones correspond to $k<l$,  co-occurring ones to $k=l$ and  enclosing ones to $k>l$. Similarly, from the disjoint contribution, mutations belonging to the strictly disjoint component can be obtained by selecting the frequency range $k+l<n$ while complementary ones correspond to $k+l=n$. 


\subsection{Population spectra}

In the limit of large samples, the frequency spectra converge to the continuous SFS for infinite populations. However, the limit $n\rightarrow \infty$ should be taken with care. The easiest derivation proceeds as follows: since the conditional 1-SFS (eq \ref{sfs_lp}) is  a single-locus spectrum, its population components can be obtained from the corresponding ones for finite samples (eq. \ref{sfs_ls}) by direct application of the equation (\ref{eqlim}). Then the population 2-SFS (eq \ref{sfs_2p}) can be reconstructed from equations (\ref{eq1p}) and (\ref{eq2p}), by multiplying by the neutral spectrum $\ev[\xi(f_0)]=\theta L/f_0$ and by $\frac{1}{1+\delta_{f,f_0}}$ and combining the result into nested and disjoint contributions.  
The only tricky passage of the derivation is the following functional limit of the Kronecker delta as a Dirac delta function: 
$n\delta_{\lfloor nf\rfloor ,\lfloor nf_0\rfloor }\rightarrow \delta(f-f_0)$ for $n\rightarrow \infty$. More details are
given in the Supplementary Material.




\section{Discussion}

In this article, we have provided the first exact closed formulae for the joint 2-SFS as well as for the conditional
1-SFS, both for sample and population. Using the basic results from \cite{fu1995statistical}, we were able to derive the
formulae for sample spectra which we used then to derive the population spectra by letting $n \to \infty$.
Importantly, our results only hold when there is no recombination, and are averaged across the tree space.

The analytical expressions provided in this paper can be intuitively understood in terms of the evolution of linked mutations. Consider a new mutation increasing in frequency by neutral drift and reaching low/intermediate frequency. We expect to find a large number of strictly disjoint and a low number of strictly nested linked mutations, since at the time of appearance of the focal mutation all other mutations were ``strictly disjoint''. Enclosing mutations are more abundant than strictly nested, but less or  abundant as strictly disjoint mutations, depending on the initial frequency of the focal mutation. The spectrum of strictly nested mutations is more skewed towards rare alleles than predicted by the neutral spectrum $1/f$, since strictly nested mutations evolve inside an expanding subpopulation. On the other hand, the spectrum of strictly disjoint mutations resembles the neutral one but with a slight bias against rare mutations, since they evolved in a slightly contracting subpopulation. 

Note that for sequences linked to a mutation close to fixation,  co-occurring and complementary mutations dominate. The contrast between the haplotypes produces a strong ``haplotype structure''.

Interestingly, conditioning on the presence of a mutation of frequency $f$ impacts the length and balance of the coalescent, as apparent from Figure \ref{fig_wattpi}. This can be understood as follows. Rare mutations are common in any realisation of the coalescent tree but especially common in the lower branches, therefore they just increase slightly the tree length and the length of the lower branches compared to the unconditioned case. Instead, mutations of intermediate frequency appear mostly in the upper branches of the tree, therefore the presence of such mutations implies higher, more balanced trees. The effect is even stronger for high frequency mutations, which reside only in the uppermost branches, implying high unbalanced trees.

There are several potential applications of these results. Direct applications include the improvement of population genetic inference techniques based on the SFS, such as composite likelihood \citep[\textit{e.g., }][]{Kim:2002xe,Li:2005zv,Kim:2004sj,Nielsen:2005bs} and Poisson Random Field methods \citep{sawyer1992population}. These methods use analytical expressions for the SFS for a single site together with approximations of independence between different sites. For sequences with low recombination, they could be made more rigorous by assuming independence between different \emph{pairs} of sites, while taking pairwise dependence between sites into account through the two-locus SFS developed here. 

The spectrum could also be useful for new neutrality tests based on linkage between mutations. Our results lead to a better understanding of the linkage disequilibrium (LD) structure among neutral loci, therefore they can be immediately applied to LD-related statistics, for example to compute average LD across non-recombining neutral loci. Furthermore, they can be used to build neutrality tests optimised to detect positive or balancing selection through its effect on the frequency spectrum of linked sites.

\commblue{An example of direct application of our results is the spectrum of chromosomal inversions and other structural variants. These genomic variants often have phenotypic effects and their evolutionary dynamics is of significant interest. We provided the spectrum of the null neutral model, that is a fundamental step to build methods for the detection of non-neutral evolution. Further work on the derivation of appropriate neutrality tests, their optimisation and application will be presented in future publications. 

The spectra presented here could also provide a neutral model for other scenarios, including introgressions from different species or populations.  Our results contribute to the ongoing search for genetic signatures of selection on introgressed alleles.}

The SFS presented here is the simplest two-locus spectrum for neutral, non-recombining mutations in a population of constant size. These results could be extended to variable population size using the approach of \cite{vzivkovic2008second,jenkins2011effect} 
and to mutations in rapidly adapting populations using the $\Lambda$-coalescent approximation and the results of \cite{birkner2013statistical}. However, the most interesting extensions would be to consider (a) non-neutral mutations and (b) recombination.

Adding selection to the two-locus SFS would significantly enhance its potential for most of the applications discussed above. The SFS for pairs of selected mutations has been obtained by \cite{xie2011site} as a polynomial expansion. However, the computation is still cumbersome, while flexible numerical alternative could be soon available. Given the simplicity of the expression for the single-locus SFS $\xi(f)=\theta(1-e^{-2N_es(1-f)})/f(1-f)(1-e^{-2N_es})$ \citep{wright1938distribution,sawyer1992population}, we expect that closed expressions could be found for pairs of mutations with different selective coefficients. This would be a promising development for future investigations.

The classical correspondence between the Kingman model in the large $n$ limit and the diffusion approximation suggests that the 2-SFS spectrum presented here is a solution of the diffusion equations for three alleles \citep{ewens2012mathematical}. 
In fact, it is easy to check that the nested component of the 2-SFS for $f\neq f_0$ is a stationary solution of the diffusion equation of three alleles of frequency $f$, $f_0-f$ and $1-f_0$:
\beq
\frac{\partial \xi}{ \partial t}=\frac{1}{2N_e}\left(
\frac{\partial^2}{\partial f^2} \left[f(1-f)\xi\right] +2 \frac{\partial^2}{\partial f \partial f_0} \left[f(1-f_0)\xi\right]+ \frac{\partial^2}{\partial f_0^2} \left[f_0(1-f_0) \xi\right]\right)
\eeq
while the disjoint component for $f\neq 1-f_0$ is a stationary solution of the diffusion equation of three alleles of frequency $f$, $f_0$ and $1-f_0-f$:
\beq
\frac{\partial \xi}{ \partial t}=
\frac{1}{2N_e}\left(
\frac{\partial^2}{\partial f^2} \left[f(1-f)\xi\right]-2\frac{\partial^2}{\partial f \partial f_0} \left[ff_0\xi\right] + \frac{\partial^2}{\partial f_0^2} \left[f_0(1-f_0)\xi\right]\right)
\eeq
The correspondence implies that the solution (\ref{sfs_2p}) is actually the stationary solution of the full set of diffusion equations for the system, including boundary equations for $f=f_0$ and $1-f_0$ and boundary conditions. A direct proof of this result using methods from the theory of partial differential equations could lead to interesting developments towards new solutions for selective equations as well.

On the other hand, finding the exact two-locus SFS with recombination appears to be a difficult problem. Recombination is intrinsically related to the two-locus SFS via the same definition of linkage disequilibrium. Obtaining the full two-locus spectrum with selection and recombination could open new avenues for model inference and analysis of genomic data. For this reason, many approximations and partial results have been developed since \cite{hudson2001two}, like expansions in the limit of strong recombination \citep{jenkins2012pade}. The SFS of linked loci presented in this paper could be useful as a starting point for different approaches to the effect of recombination events, for example for perturbation expansions at low recombination rates.

There is actually an immediate application of our results to recombination events. Since in the Ancestral Recombination Graph \citep{griffiths1997ancestral} the recombination events follow a Poisson process similar to mutation events, although with a different rate, the spectrum $\xi_{k|l}$ could also be reinterpreted (up to a constant) as the probability that a single \emph{recombination} event affects $k$ extant lineages in a sequence linked to a specific mutation of frequency $l$, i.e. it is equivalent to the spectrum of mutation-recombination events. This approach could be applied to higher moments of the frequency spectrum and lead to new results in recombination theory.

\section*{Acknowledgments}
We thank 
Wolfgang Stephan for insightful discussions. 
GA and LF were supported by grant ANR-12-JSV7-0007 TempoMut from Agence Nationale de la Recherche. GA was
also supported by grant ANR-12-BSV7-0012 Demochips, AK and TW by grants of the German Science Foundation (DFG-SFB680 and DFG-SPP 1590).

\bibliographystyle{genetics}
\bibliography{fs-linked}


%
%

\clearpage

\clearpage

\appendix

\section{2-SFS for ordered pairs of sites}\label{app_ordered}

The expected spectrum of linked sites described in the previous sections applies to unordered pairs of sites. As an
example, consider a sequence containing just two nested SNPs with mutations of frequency
0.3 and 0.1 respectively. The nonzero components of the spectrum are $\xi(0.1,0.3)=\xi(0.3,0.1)=1$, irrespective of
which of the two SNPs has frequency 0.1.

However, it can be useful to rewrite our results in terms of the spectrum $\xi^{ordered}$ for ordered pairs of sites.
Sites can be ordered by their position along the sequence, or by any other criterion. In the previous example, the
components of the ordered spectrum are $\xi^{ordered}(0.3,0.1)=1$ but $\xi^{ordered}(0.1,0.3)=0$.

The relation between the 2-SFS and the ordered 2-SFS is the following. For different frequencies $k\neq l$, the 2-SFS of unordered pairs is symmetric, so $\xi_{k,l}=\xi_{l,k}$ are actually the same
object. However, for the ordered 2-SFS, they are different. Their sum correspond to the total number of unordered pairs:
\beq
\xi_{k,l} = \xi_{k,l}^{ordered} + \xi_{l,k}^{ordered}
\eeq
and since the expected values do not depend on the order,
\beq
\ev[\xi_{k,l}^{ordered}]=\ev[\xi_{l,k}^{ordered}]=\ev[\xi_{k,l}]/2
\eeq
On the other hand, the order does not matter for pairs of identical mutations, i.e. $\xi_{k,k} = \xi_{k,k}^{ordered}$ and therefore
\beq
 \ev[\xi_{k,k}^{ordered}]=\ev[\xi_{k,k}]
 \eeq
These relations can be extended to the population spectrum in a straightforward way. Note that this factor of 2 between both cases relates to the same factor in equations (\ref{eq_spec_prob_p}) and (\ref{eq_spec_prob_s}). In fact, $\ev[\xi_{k,l}^{ordered}]=\ev[\xi_{k|l}]\cdot \ev[\xi_l]$.

\section{Triallelic spectrum}\label{app_triallelic}

As discussed before, the evolutionary dynamics of two non-recombining SNPs is the same as the one of a triallelic locus, where the three alleles are represented by the possible haplotypes of the sequence containing the SNPs. Therefore we can extract the frequency spectrum of neutral mutations in a triallelic non-recombining locus from our results. 

Triallelic loci can represent many possible types of variants in genomes. Triallelic SNPs can be present in any set of nucleotide sequences - however these sites are rare compared to biallelic SNPs. Or they could be Copy Number Variants, or microsatellites with variable number of repeats.


The unfolded tri-allelic spectrum for two derived alleles of frequency $f_1, f_2$ generated with rescaled mutation rates per locus $\theta_1^{loc},\theta_2^{loc}$ is
\beq
\ev[\xi^{3al}(f_1,f_2)]={\theta_1\theta_2}\left(\ev[\xi^{D}(f_1,f_2)]+\ev[\xi^{N}(f_2-f_1,f_1)]+\ev[\xi^{N}(f_1-f_2,f_2)]\right),
\eeq
where the expectations are given by equation (\ref{sfs_2p}) with $\theta=1$. 


Similarly, the sample triallelic spectrum for derived alleles of count $k,l$ is
\beq
\ev[\xi^{3al}_{k,l}]={\theta_1\theta_2}\left(\ev[\xi^{D}_{k,l}]+\ev[\xi^{N}_{k-l,l}]+\ev[\xi^{N}_{l-k,k}]\right),
\eeq
where the expectations are given by equation (\ref{sfs_2s}) with $\theta=1$. This spectrum was also derived by \citet{jenkins2011effect} for a general matrix of mutation rates.

\section{The folded spectra}\label{app_folded}

When no reliable outgroup sequence is available, one cannot assess if the allele is derived or ancestral. In that case, alleles can only be classified as minor (less frequent) and major (most frequent).  The distribution of minor allele frequencies, known as the folded SFS, will be noted $\eta(f^*)$, where $f^*$ denotes the minor allele frequency that ranges from 0 to 0.5. Importantly, the folded SFS can be retrieved from the full SFS by simply summing alleles at complementary frequencies:

\beq
	\eta(f^*) = [ \xi(f^*) + \xi(1-f^*) ] / (1+\delta_{f^*,(1-f^*)})
\eeq

As a consequence, the single site SFS under the standard neutral model then become $\ev[\eta(f^*)]=\theta/[ f^*(1-f^*)(1+\delta_{f^*,(1-f^*)}) ]$ and $\ev[\eta_{k^*}]=\theta n/[ k^*(n-k^*)(1+\delta_{k^*,n-k^*}) ]$, where $k^*$ denotes the count of the minor allele.


Following the same idea, we define a  conditional folded 1-SFS and a joint folded 2-SFS using the minor allele frequencies.  Minor alleles can also be classified as ``nested'' or ``disjoint'' depending on the presence or absence of individuals enclosing both minor alleles. As for the unfolded case, this classification gives a complete description of the linkage between pairs of mutations. However, in contrast to the unfolded case, the classification has no strict evolutionary meaning. For example, ``disjoint'' minor alleles do not necessarily correspond to pairs of alleles born in different backgrounds. Moreover, alleles of frequency $f^*=0.5$ (or allele count $k^*=n/2$) suffer from an ambiguity in the choice of the minor allele and therefore should be treated separately.  Note also that with the exception of alleles with frequency 0.5, folded spectra do not contain complementary alleles, since the frequency of one of the two complementary alleles will exceed 0.5. 

Pairs of mutations with $f,f_0$ both larger or smaller than $0.5$ will be classified identically (as nested or disjoint) in the folded case. However, pairs of mutations with $f<0.5$ and $f_0>0.5$ (or vice-versa) will swap their classification. As a consequence, the two components of the 2-SFS are:
\begin{align}
\ev[\eta^{N}(f^*,f^*_0)]=& \ev[\xi^{N}(f^*,f^*_0)]+\ev[\xi^{N}(1-f^*,1-f^*_0)] + \ev[\xi^{D}(f^*,1-f^*_0)] \nonumber \\\
		&+\ev[\xi^{D}(1-f^*,f^*_0)] \nonumber \\
\ev[\eta^{D}(f^*,f^*_0)]=& \ev[\xi^{D}(f^*,f^*_0)]+\ev[\xi^{N}(f^*,1-f^*_0)]+\ev[\xi^{N}(1-f^*,f^*_0)] 
\end{align}

 To obtain the conditional 1-SFS, we proceed similarly to the unfolded case. First we separate the 2-SFS above into components based on frequency. The strictly nested component corresponds to frequencies $f^*<f^*_0$ of the nested part, while the cooccurring and enclosing components corresponds to $f^*=f^*_0$ and  $f^*>f^*_0$ respectively. The strictly disjoint component corresponds to the disjoint part, since there cannot be any complementary component. Then we divide each component by the expected 1-SFS $\ev[\eta(f^*_0)]$ to obtain
\begin{align}
\ev[\eta^{(sn)}(f^*|f^*_0)]=& \frac{f^*_0(1-f^*_0)}{\theta} \ev[\eta^{N}(f^*,f^*_0)] \quad \mathrm{for}\ f^*<f^*_0 \nonumber\\
\ev[\eta^{(co)}(f^*|f^*_0)]=&2\cdot \frac{f^*_0(1-f^*_0)}{\theta} \ev[\eta^{N}(f^*,f^*_0)] \quad \mathrm{for}\ f^*=f^*_0 \nonumber\\
\ev[\eta^{(en)}(f^*|f^*_0)]=& \frac{f^*_0(1-f^*_0)}{\theta} \ev[\eta^{N}(f^*,f^*_0)] \quad \mathrm{for}\ f^*>f^*_0 \\
\ev[\eta^{(cm)}(f^*|f^*_0)]=& 0\nonumber\\
\ev[\eta^{(sd)}(f^*|f^*_0)]=& (1+\delta_{f^*,f^*_0}) \cdot\frac{f^*_0(1-f^*_0)}{\theta}\ev[\eta^{D}(f^*,f^*_0)]\nonumber
\end{align}

While the classification of the pairs with frequencies $f^*=0.5$ and/or $f^*_0=0.5$ is ambiguous, these pairs are usually irrelevant for the population spectrum.

The sample spectra are similar. For $n$ even, there are ambiguous pairs with $k$ or $l=n/2$ that can be easily retrieved from the equations (\ref{sfs_2s}),(\ref{sfs_ls}) and treated separately. Considering only $k,l<n/2$, the sample 2-SFS is:
\begin{align}
\ev[\eta^{N}_{k^*,l^*}]=& \ev[\xi^{N}_{k^*,l^*}]+\ev[\xi^{N}_{n-k^*,n-l^*}] + \ev[\xi^{D}_{k^*,n-l^*}]+\ev[\xi^{D}_{n-k^*,l^*}] \nonumber \\
\ev[\eta^{D}_{k^*,l^*}]=& \ev[\xi^{D}_{k^*,l^*}]+\ev[\xi^{N}_{k^*,n-l^*}]+\ev[\xi^{N}_{n-k^*,l^*}] 
\end{align}
and the conditional 1-SFS is:
\begin{align}
\ev[\eta^{(sn)}_{k^*|l^*}]=& \frac{l^*(n-l^*)}{\theta n} \ev[\eta^{N}_{k^*,l^*}] \quad \mathrm{for}\ k^*<l^* \nonumber \\
\ev[\eta^{(co)}_{k^*|l^*}]=& 2\cdot\frac{l^*(n-l^*)}{\theta n} \ev[\eta^{N}_{k^*,l^*}] \quad \mathrm{for}\ k^*=l^* \nonumber \\
\ev[\eta^{(en)}_{k^*|l^*}]=& \frac{l^*(n-l^*)}{\theta n} \ev[\eta^{N}_{k^*,l^*}] \quad \mathrm{for}\ k^*>l^* \\
\ev[\eta^{(cm)}_{k^*|l^*}]=& 0\nonumber \\
\ev[\eta^{(sd)}_{k^*|l^*}]=&(1+\delta_{k^*,l^*})\cdot \frac{l^*(n-l^*)}{\theta n} \ev[\eta^{D}_{k^*,l^*}]\nonumber
\end{align}

%
%

\clearpage

\section*{Supplementary Material}

\renewcommand\thefigure{S\arabic{figure}}    

\setcounter{section}{19}

\setcounter{figure}{0}    

\subsection{Classification of two linked mutations}


As discussed also by \cite{Sargsyan:2015hw}, it is easy to see that our mutation classes cover all possible
relations of two mutations in a non-recombining coalescent. The two bi-allelic sites were created by two independent mutations: an {\em old} mutation followed by a {\em young} one. They both occurred in a single individual and then rose in frequency throughout the action of genetic drift. The young mutation could have occurred in an individual that also carried the old mutation, leading to what we have name the ``nested'' case. 

Conversely, if the young mutation has occurred in an individual who did not have the old mutation, it leads to the ``disjoint'' case. As recombination is forbidden here, the complete linkage prevents any further mixing between these two cases and the derived allele that corresponds to the young mutation will remain fully linked to the background allele it occurred in.

In the nested case, the young mutation can be fixed in sequences carrying the old one (that is co-occurring case) or not. In the latter case, the young mutation can be the focal one (enclosing case) or the other one (strictly nested case).
 
In the disjoint case, the young mutation can get fixed among the individuals lacking the old mutation (complementary case) or not (strictly disjoint case). Therefore, without recombination, these 5 types are the only possible cases.

Because these are the only possible classes of mutations without recombination, there are constrains on the frequency spectrum for linked sites. For example, the presence of an enclosing mutation of count $k$ is incompatible with complementary mutations or strictly disjoint mutations of count greater than $n-k$; this can be shown by considering the enclosing mutation as focal one, and noticing that the other mutations would not fall in any of the previous classes.

\subsection{Simulations}

\begin{figure}[h!]
\begin{center}
\includegraphics[width=12cm]{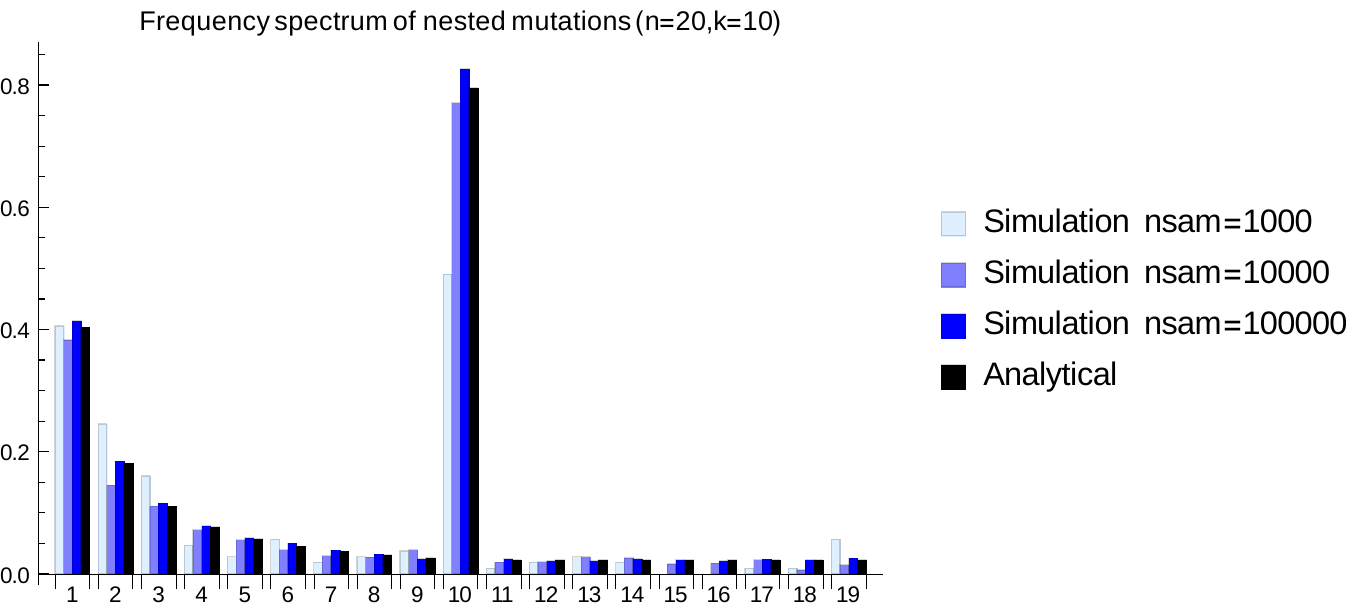}
\caption{Frequency spectrum of nested mutations in linked sites for n=20, $L\theta=1$ and a focal mutation of frequency
$k=10$, compared with coalescent simulations (averages for different numbers of samples).}\label{fig_sfssim}
\end{center}
\end{figure}

In this section we present a numerical result as an example to check the consistency of our results.
In Figure \ref{fig_sfssim} the analytical sample spectrum is compared with those obtained by coalescent simulations. 
We parsed the output of \emph{ms} \citep{Hudson:2002td} to count the number of mutations conditional on a focal mutation
of given frequency. The good agreement between the spectra supports our equations. 

The source code (C++) for computing
analytical as well as simulated spectra can be found in the package \emph{coatli} developed by one of the authors and available on \url{http://sourceforge.net/projects/coatli/}.

\subsection{Derivation of the sample 2-SFS: \cite{fu1995statistical} reloaded}

The 1995 paper by \cite{fu1995statistical} derived the second moments of the Kingman coalescent
\citep{kingman1982coalescent}, more precisely the covariance of mutations of size $i$ and $j$: Cov[$\xi_i,\xi_j$].
Unfortunately the very tight presentation and some typos may make it
hard to follow the transformations. A valuable introduction into the proof, using a different notation, has been
given in \citet{Durrett2008}, omitting the more technical parts. Since the latter are important for us, we reproduce
the essential parts of the proof in greater detail and original notation, and show how they lead to our
expressions for different mutations classes.\\[2ex]

\begin{figure}[h!]
\begin{center}
\includegraphics[width=10cm]{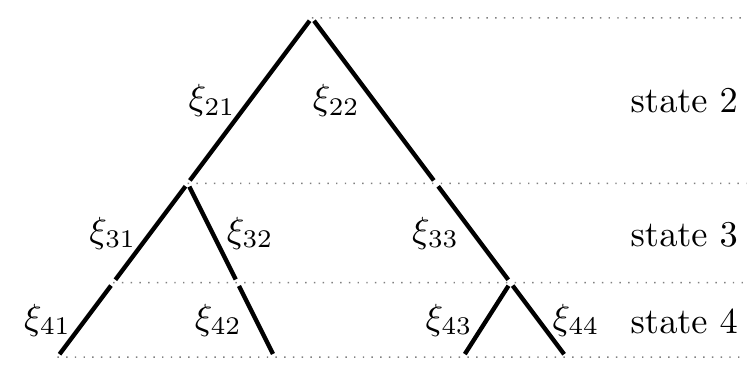}
\caption{A coalescent tree describing the genealogy of a non-recombining locus for a sample of size $n=4$. The
topology of the tree is defined by the relationship between the lines, e.g. line $\xi_{43}$ is a descendant of lines
$\xi_{33}$ and $\xi_{22}$, but not of any other line.
A mutation happening ``on'' line $\xi_{33}$ is of size $2$, since it has two descendant lines (and hence leaves) at
state $n=4$, i.e.
two individuals of the sample carry it. All lines of the same state have the same length, reflecting
the same mutation probability. Hence the amount of mutations of size $1$ (``singletons'') occurring on $\xi_{31}$
and $\xi_{32}$ is correlated with the amount of mutations of size $2$ arising on $\xi_{33}$.
Averaging over different topologies leads to more complicated correlations.}\label{fig_Fu1}
\end{center}
\end{figure}

As a starting point for the combinatorics let us note that the descendance of lines in the coalescent can be
described by a Polya urn process, and the two expressions given beneath are special cases of a general formula (c.f.
e.g.
\cite{griffiths2003genealogy}). We introduce the following notation:
let $p_{k\sra n}(t\sra i)$ denote the probability that $t$ lines at state $k$ have $i$ descendents at state $n$.
This probability is
 \begin{align*}
 p_{k\sra n}(t\sra i)=
 \frac{{i-1\choose t-1}{n-i-1\choose k-t-1}}{{n-1\choose k-1}}
 \end{align*}
 and the probability that $t$ and $u$ lines at state $k$ have respectively
 $i$ and $j$ descendents at state $n$ is
 \begin{align*}
 p_{k\sra n}(t\sra i,u\sra j)=\frac{{i-1\choose t-1}{j-1\choose u-1}{n-i-j-1\choose k-t-u-1}}{{n-1\choose k-1}}\;.
 \end{align*}
 In order to avoid case distinctions it is helpful to abuse for a while the notation by defining
 ${{-1\choose -1}}=1$ and ${{n \choose k}}=0$ for any other combination of $n<0$ or $k<0$ (as has been employed by
 \citet{Durrett2008}, too).
 This makes it possible to subsume in the above and following formulas ``boundary cases'' such as $k$ lines of
 state $k$ yielding the $n$ lines of state $n$ (with probability $1$). Later on these special cases
 will be considered separately and the final expressions don't contain any negative values.
 
  The probability that a line at state $k$ is of size $i$ is referred to as $p(k,i)$. The probability that two lines at
 state $k$ are of size $i$ and $j$ is referred to as $p(k,i;k,j)$. The probability that a line at state $k$ and another
 at state $k'>k$ are of size $i$ respective $j$ is split up with respect to the latter line
 being a descendant of the former line or not: $p(k,i;k',j)=p_a(k,i;k',j)+p_b(k,i;k',j)$. \\
 
 The two formulas above suffice to derive these probabilities:
\begin{align*}
p(k,i)&=p_{k\sra n}(1\sra i)=\frac{{n-i-1\choose k-2}}{{n-1\choose k-1}}\\
p(k,i;k,j)&=p_{k\sra n}(1\sra i,1\sra j)=\frac{{n-i-j-1\choose k-3}}{{n-1\choose k-1}}\\
p_a(k,i;k',j)&=\sum_{t=1}^{k'-1} p_{k\sra k'}(1\sra t)\frac{t}{k'}p_{k'\sra n}(1\sra j,t-1 \sra i-j)\\
&=\sum_{t=1}^{k'-1}\frac{{k'-t-1\choose
k-2}}{{k'-1\choose k-1}}\frac{t}{k'}\frac{{i-j-1\choose t-2}{n-i-1\choose k'-t-1}}{{n-1\choose k'-1}}\\
p_b(k,i;k',j)&=\sum_{t=1}^{k'-1} p_{k\sra k'}(1\sra t)\frac{k'-t}{k'}p_{k'\sra n}(1\sra j,t\sra i)\\
&=\sum_{t=1}^{k'-1} \frac{{k'-t-1\choose
k-2}}{{k'-1\choose k-1}}\frac{k'-t}{k'}\frac{{i-1\choose t-1}{n-i-j-1\choose k'-t-2}}{{n-1\choose k'-1}}
\end{align*}
In the latter two formulas, the summation index $t$ stands for the number of descendants, that the line from state
$k$ may have at state $k'$.\\

Now we consider the ``mutational'' correlation between the lines. Other than in our main article, $\theta$ denotes
here the {\em locus} mutation rate (not the site mutation rate), i.e. includes the locus length $L$.\\

Let $X$ be a random variable. It can be easily shown that, if $X$ is exponentially distributed ($X\sim Exp(\lambda)$),
then the first two moments of $X$ are $E[X]=\frac{1}{\lambda}$ and $E[X^2]=\frac{2}{\lambda^2}$. If $X$ is
Poisson-distributed ($X\sim Poiss(\mu)$), then $E[X]=\mu$ and $E[X^2]=\mu+\mu^2$.
By definition of the coalescent the $\xi_{kl}$ are distributed like $\xi_{kl}\sim
Poiss(\frac{\theta}{2}T_k)$ with $T_k\sim Exp(\frac{2}{k(k-1)})$. $\xi_{kl}$ and $\xi_{k'l'}$ are independent if
$k\neq k'$ while $\xi_{kl}$ and $\xi_{kl'}$ are independent conditional on $T_k$ for $l\neq l'$. We have thus
$$T_k\sim Exp(\lambda_k) \text{ with } \lambda_k=\frac{k(k-1)}{2} \text{ and }\xi_{kl}\sim
Poiss(\mu)\text{ with }\mu=\frac{\theta}{2}T_k$$
\begin{align*}
E[\xi_{kl}]&=E[E[\xi_{kl}|T_k]]=E[\frac{\theta}{2}T_k]=\frac{1}{k(k-1)}\theta\\[3ex]
E[\xi_{kl}^2]&=E[E[\xi_{kl}^2|T_k]]\\
&=E[\frac{\theta}{2}T_k+(\frac{\theta}{2}T_k)^2]\\
&=\frac{\theta}{2}E[T_k]+\frac{\theta^2}{4}E[T_k^2]\\
&=\frac{2}{k(k-1)}\frac{\theta}{2}+2\frac{2^2}{k^2(k-1)^2}\frac{\theta^2}{4}\\
&=\frac{1}{k(k-1)}\theta+\frac{2}{k^2(k-1)^2}\theta^2\\[3ex]
E[\xi_{kl}\xi_{kl'}]=&E[E[\xi_{kl}\xi_{kl'}|T_k]]\\
&=E[E[\xi_{kl}|T_k]E[\xi_{kl'}|T_k]]\\
&=E[(\frac{\theta}{2}T_k)^2]\\
&=\frac{2}{k^2(k-1)^2}\theta^2\\[3ex]
E[\xi_{kl}\xi_{k'l}]&=E[\xi_{kl}]E[\xi_{k'l}]\\
&=\frac{1}{k(k-1)k'(k'-1)}\theta^2\\
\end{align*}
For a particular topology, the number of mutations of size $i$ can be parcelled onto lines as
$$\xi_i=\sum_{k=2}^n\sum_{l=1}^k\epsilon_{kl}(i)\xi_{kl}$$
with the ``indicator-variable'' $\epsilon_{kl}(i)=1$ if line $\xi_{kl}$ has $i$ descendent leaves and $0$
otherwise. We take the expectation over all topologies and branch lengths:
\begin{align*}
E[\xi_i\xi_j]=&E[(\sum_{k=2}^n\sum_{l=1}^k\epsilon_{kl}(i)\xi_{kl})(\sum_{k'=2}^n\sum_{l'=1}^{k'}\epsilon_{k'l'}(j)\xi_{k'l'})]\\
=&\sum_{k=2}^n\sum_{k'=2}^n\sum_{l=1}^k\sum_{l'=1}^{k'}E[\epsilon_{kl}(i)\epsilon_{k'l'}(j)]E[\xi_{kl}\xi_{k'l'}]\\
=&\sum_{k=2}^n\sum_{l=1}^kE[\epsilon_{kl}(i)\epsilon_{kl}(j)]E[\xi_{kl}\xi_{kl}]+\sum_{k=2}^{n}\sum_{l=1}^{n-1}\sum_{l'=l+1}^nE[\epsilon_{kl}(i)\epsilon_{kl'}(j)]E[\xi_{kl}\xi_{kl'}]+\\
&\sum_{k=2}^{n-1}\sum_{k'=k+1}^n\sum_{l=1}^k\sum_{l'=1}^{k'}\left(E[\epsilon_{kl}(i)\epsilon_{k'l'}(j)+E[\epsilon_{k'l}(i)\epsilon_{kl'}(j)]\right)E[\xi_{kl}\xi_{k'l'}]\\
=&\delta_{i=j}\sum_{k=2}^nkp(k,i)E[\xi_{kl}^2]+\sum_{k=2}k(k-1)p(k,i;k,j)E[\xi_{k1}\xi_{k2}]\\
&+\sum_{k=2}^{n-1}\sum_{k'=k+1}^nkk'(p(k,i;k',j)+p(k,j;k',i))E[\xi_{k1}\xi_{k'1}]
\end{align*}\\

If we define for $k<k'$
\begin{align*}
s_1(i)&=\sum_{k=2}^nkp(k,i)\frac{1}{k(k-1)}\\
s_2(i)&=\sum_{k=2}^nkp(k,i)\frac{2}{k^2(k-1)^2}\\
s(i,j)&=\sum_{k=2}^nk(k-1)p(k,i;k,j)\frac{2}{k^2(k-1)^2}\\
s_a(i,j)&=\sum_{k=2}^{n-1}\sum_{k'=k+1}^nkk'p_a(k,i;k',j)\frac{1}{k(k-1)k'(k'-1)}\\
s_b(i,j)&=\sum_{k=2}^{n-1}\sum_{k'=k+1}^nkk'p_b(k,i;k',j)\frac{1}{k(k-1)k'(k'-1)}
\end{align*}
then
\begin{align*}
E[\xi_i\xi_j]&=\delta_{i=j}s_1(i)\,\theta\;+\\
&\quad\left(\delta_{i=j}s_2(i)+s(i,j)+s_a(i,j)+s_a(j,i)+s_b(i,j)+s_b(j,i)\right)\theta^2\;.
\end{align*}
The different relations between lines correspond to our subdivision of the conditional frequency spectrum. In
particular, we have
\begin{align*}
\ev[\xi^{(sn)}_{i|j}]=&\delta_{i<j}\theta^2j\,s_a(j,i)\\
\ev[\xi^{(co)}_{i|j}]=&\delta_{i=j}\theta^2j\,\left(s_2(i)+2s_a(i,i)\right)\\
\ev[\xi^{(en)}_{i|j}]=&\delta_{i>j}\theta^2j\,s_a(i,j)\\
\ev[\xi^{(cm)}_{i|j}]=&\delta_{i+j=n}\theta^2j\left(s(i,j)+s_b(i,j)+s_b(j,i)\right)\\
\ev[\xi^{(sd)}_{i|j}]=&\delta_{i+j<n}\theta^2j\left(s(i,j)+s_b(i,j)+s_b(j,i)\right).
\end{align*}
 
The following derivations simplify these expressions until we finally yield the equations \ref{sfs_ls}. 

The simplification makes use of two known formulas for binomial coefficients:
\begin{align}
\sum_{m=0}^n{m\choose k}={n+1\choose k+1}\label{eq:bi1}\tag{B1}
\end{align}
\begin{align}
\sum_{j=0}^k{m\choose j}{n-m\choose k-j }={n\choose k}\label{eq:bi2}\tag{B2}
\end{align}
In the first equation, the summation can
start as well at $m=k$ since ${m\choose k}=0$ for $m<k$.\\[3ex]

Furthermore we need three helping equations from \cite{fu1995statistical}:\\
The straight-forward computable equation (14)
\begin{align*}
\frac{{n-i-1\choose k-2}}{{n-1\choose k-1}}=\frac{{n-k \choose i-1}}{{n-1\choose i}}\frac{k-1}{i} 
\end{align*}
and the technically more demanding equations (34) 
\begin{align*}
2\sum_{k=2}^n\frac{{n-k\choose i-1}}{{n-1\choose i}i}\frac{1}{k}=\beta_n(i)
\end{align*}
and (36)
\begin{align*}
\sum_{k=3}^n\frac{{n-i-2\choose k-3}}{{n-1\choose k-1}}\frac{1}{k(k-1)}=\frac{\beta_n(i)-\beta_n(i+1)}{2}\quad.
\end{align*}
A useful variation of equation (34), needed repeatedly, can be derived using his equation (33) (not replicated here):
\begin{align*}
\frac{1}{{n-1\choose i}i}\sum_{k=2}^n \frac{{n-k \choose i-1}}{k-1}
&=\frac{1}{{n-1\choose i}i}\sum_{k=1}^{n-1}\frac{{n-1-k\choose i-1}}{k}\nonumber\\
&\overset{\mathclap{(33)}}=\frac{1}{n-i}\frac{1}{{n-1\choose i-1}}{{n-1\choose i-1}}(a_n-a_i)\nonumber\\
&=\frac{a_n-a_i}{n-i}\quad.\label{eq:34a}\tag{34a}\\[3ex]
\end{align*}
Now we have to account for the ``boundary cases'' in the probability expressions $p()$. As defined above, a binomial
coefficient ${a \choose b}$ with $a=-1$ is non-zero only for $b=-1$, which translates to
additional constraints on the state $k$ and the number of descendants that the line from this state can have at state $k'$:\\
For example, if in the expression
 \begin{align*}
 p(k,i;k,j)=\frac{{n-i-j-1\choose k-3}}{{n-1\choose k-1}}
 \end{align*}
we have $i+j=n$, then the descendants of two lines encompass the whole sample. However this is only possible for
the two lines of state $k=2$. 
The same reasoning applied on $p_a(k,i;k',j)$ for $i=j$ leads to the condition, that the summation is only over one
element, namely $t=1$.
Finally, if $i+j=n$ in the expression for $p_b(k,i;k',j)$, then $k=2$ and $t=k'-1$.
\newpage
\begingroup
\allowdisplaybreaks
\begin{align*}
s_1(i)&=\sum_{k=2}^n k p(k,i)\frac{1}{k(k-1)}\\
&\overset{\mathclap{(14)}}=\sum_{k=2}^n k 
\frac{{n-k \choose i-1}(k-1)}{{n-1 \choose
i}i}\frac{1}{k(k-1)}\\
&=\frac{1}{i}\sum_{k=0}^{n-2}\frac{{k\choose i-1}}{{n-1\choose i}}\\
&\overset{\mathclap{(B1)}}=\frac{1}{i}\\[4ex]
s_2(i)&=\sum_{k=2}^n k p(k,i)\frac{2}{k^2(k-1)^2}\\
&\overset{\mathclap{(14)}}=\sum_{k=2}^n k 
\frac{{n-k \choose i-1}(k-1)}{{n-1 \choose
i}i}\frac{2}{k^2(k-1)^2}\\
&=\sum_{k=2}^n\frac{{n-k \choose i-1}}{{n-1
\choose i}i}\frac{2}{k(k-1)}\\
&=2\sum_{k=2}^n \frac{{n-k \choose i-1}}{{n-1  \choose i}i}(\frac{1}{k-1}-\frac{1}{k})\\
&\overset{\mathclap{(34),(\ref{eq:34a})}}=\quad2\frac{a_n-a_i}{n-i}-\beta_n(i)\\[4ex]
s(i,j)&=\sum_{k=2}^n k(k-1)p(k,i;k,j)\frac{2}{k^2(k-1)^2}\\
&=\sum_{k=2}^n \frac{{n-i-j-1 \choose
k-3}}{{n-1 \choose k-1}}\frac{2}{k(k-1)}\\
&=\delta_{i+j<n}\sum_{k=3}^n \frac{{n-(i-j-1)-2 \choose
k-3}}{{n-1 \choose k-1}}\frac{2}{k(k-1)}+\delta_{i+j=n}\frac{1}{n-1}\\
&\overset{\mathclap{(36)}}=\delta_{i+j<n}\left(\beta_n(i+j-1)-\beta_n(i+j)\right)+\delta_{i+j=n}\frac{1}{n-1}\\[2ex]
\intertext{Case $i>j$ ($\Rightarrow$ $t\ge 2$)}
s_a(i,j)&=\sum_{k'=3}^n\sum_{k=2}^{k'-1}kk'p_a(k,i;k',j)\frac{1}{k(k-1)k'(k'-1)}\\
&=\sum_{k'=3}^n\sum_{k=2}^{k'-1}\sum_{t=2}^{k'-1}\frac{{k'-t-1
\choose k-2}}{{k'-1 \choose k-1}}\frac{t}{k'}\frac{{i-j-1 \choose t-2}{n-i-1 \choose k'-t-1}}{{n-1
\choose k'-1}}\frac{1}{(k-1)(k'-1)}\\
&\overset{\mathclap{(14)}}=\sum_{k'=3}^n\sum_{k=2}^{k'-1}\sum_{t=2}^{k'-1}\frac{{k'-k
\choose t-1}}{{k'-1 \choose t}}\frac{{i-j-1 \choose t-2}{n-i-1 \choose k'-t-1}}{{n-1 \choose k'-1}}\frac{1}{k'(k'-1)}\\
&=\sum_{k'=3}^n\sum_{t=2}^{k'-1}\frac{{i-j-1 \choose
t-2}{n-j-2-(i-j-1) \choose k'-3-(t-2)}}{{n-1 \choose k'-1}}\frac{1}{k'(k'-1)}\sum_{k=1}^{k'-2}\frac{{k \choose
t-1}}{{k'-1 \choose t}}\\
&\overset{\mathclap{(\ref{eq:bi1})}}=\sum_{k'=3}^n\sum_{t=0}^{k'-3}\frac{{i-j-1 \choose
t}{n-j-2-(i-j-1) \choose k'-3-t}}{{n-1 \choose k'-1}}\frac{1}{k'(k'-1)}\\
&\overset{\mathclap{(\ref{eq:bi2})}}=\sum_{k'=3}^n\frac{{n-j-2 \choose k'-3}}{{n-1\choose
k'-1}}\frac{1}{k'(k'-1)}\\
&\overset{\mathclap{(36)}}=\frac{\beta_n(j)-\beta_n(j+1)}{2}
\intertext{Case $i=j$ ($\Rightarrow$ $t=1$)}
s_a(i,i)&=\sum_{k'=3}^n\sum_{k=2}^{k'-1}kk'p_a(k,i;k',i)\frac{1}{k(k-1)k'(k'-1)}\\
&=\sum_{k'=3}^n\sum_{k=2}^{k'-1}\frac{{k'-2\choose k-2}}{{k'-1\choose k-1}}\frac{1}{k'}\frac{{n-i-1 \choose
k'-2}}{{n-1\choose k'-1}}\frac{1}{(k-1)(k'-1)}\\
&=\sum_{k'=3}^n\sum_{k=2}^{k'-1}\frac{k-1}{k'-1}\frac{1}{k'}\frac{{n-i-1 \choose
k'-2}}{{n-1\choose k'-1}}\frac{1}{(k-1)(k'-1)}\\
&=\sum_{k'=3}^n\frac{{n-i-1 \choose
k'-2}}{{n-1\choose k'-1}}\frac{k'-2}{k'(k'-1)^2}\\
&\overset{\mathclap{(14)}}=\sum_{k'=3}^n\frac{{n-k' \choose
i-1}}{{n-1\choose i}i}\frac{k'-2}{k'(k'-1)}\\
&=\sum_{k'=2}^n\frac{{n-k'\choose
i-1}}{{n-1\choose i}i}\left(\frac{2}{k'}-\frac{1}{k'-1}\right)\\
&\overset{\mathclap{(34),(\ref{eq:34a})}}=\quad\beta_n(i)-\frac{a_n-a_i}{n-i}
\intertext{Case $i+j<n$ ($\Rightarrow$ $t\le k'-2$)}
s_b(i,j)&=\sum_{k'=3}^n\sum_{k=2}^{k'-1}kk'p_b(k,i;k',j)\frac{1}{k(k-1)k'(k'-1)}\\
&=\sum_{k'=3}^n\sum_{k=2}^{k'-1}\sum_{t=1}^{k'-2}\frac{{k'-t-1\choose
k-2}}{{k'-1\choose
k-1}}\frac{k'-t}{k'}\frac{{i-1\choose
t-1}{n-i-j-1\choose k'-t-2}}{{n-1\choose k'-1}}\frac{1}{(k-1)(k'-1)}\\
&\overset{\mathclap{(14)}}=\sum_{k'=3}^n\sum_{k=2}^{k'-1}\sum_{t=1}^{k'-2}\frac{{k'-k\choose
t-1}}{{k'-1\choose
t}}\frac{k'-t}{tk'}\frac{{i-1\choose
t-1}{n-i-j-1\choose k'-t-2}}{{n-1\choose k'-1}}\frac{1}{k'-1}\\
&=\sum_{k'=3}^n\sum_{k=2}^{k'-1}\left(\sum_{t=2}^{k'-2}\frac{{k'-k\choose
t-1}}{{k'-1\choose t}}\frac{k'-t}{tk'(k'-1)}\frac{{i-1\choose t-1}{n-i-j-1\choose k'-t-2}}{{n-1\choose
k'-1}}+\frac{1}{k'(k'-1)}\frac{{n-i-j-1\choose k'-3}}{{n-1\choose k'-1}}\right)\\
&=\sum_{k'=3}^n\left(\sum_{t=2}^{k'-2}\frac{k'-t}{tk'(k'-1)}\frac{{i-1\choose t-1}{n-i-j-1\choose
k'-t-2}}{{n-1\choose k'-1}}\sum_{k=1}^{k'-2}\frac{{k\choose
t-1}}{{k'-1\choose t}}+\sum_{k=2}^{k'-1}\frac{1}{k'(k'-1)}\frac{{n-i-j-1\choose k'-3}}{{n-1\choose k'-1}}\right)\\
&\overset{\mathclap{(\ref{eq:bi1})}}=\sum_{k'=3}^n\left(\sum_{t=2}^{k'-2}\frac{k'-t}{tk'(k'-1)}\frac{{i-1\choose
t-1}{n-i-j-1\choose k'-t-2}}{{n-1\choose k'-1}}+\frac{k'-2}{k'(k'-1)}\frac{{n-i-j-1\choose k'-3}}{{n-1\choose k'-1}}\right)\\
&=\sum_{k'=3}^n\sum_{t=2}^{k'-2}\frac{1}{t(k'-1)}\frac{{i-1\choose t-1}{n-i-j-1\choose
k'-t-2}}{{n-1\choose k'-1}}-\sum_{k'=3}^n\sum_{t=2}^{k'-2}\frac{1}{k'(k'-1)}\frac{{i-1\choose t-1}{n-i-j-1\choose
k'-t-2}}{{n-1\choose k'-1}}\\
&\quad+\sum_{k'=3}^n\frac{1}{k'-1}\frac{{n-i-j-1\choose k'-3}}{{n-1\choose
k'-1}}-2\sum_{k'=3}^n\frac{1}{k'(k'-1)}\frac{{n-i-j-1\choose k'-3}}{{n-1\choose k'-1}}\\
&=\sum_{k'=3}^n\sum_{t=1}^{k'-2}\frac{1}{t(k'-1)}\frac{{i-1\choose t-1}{n-i-j-1\choose
k'-t-2}}{{n-1\choose k'-1}}-\sum_{k'=3}^n\sum_{t=1}^{k'-2}\frac{1}{k'(k'-1)}\frac{{i-1\choose t-1}{n-i-j-1\choose
k'-t-2}}{{n-1\choose k'-1}}\\
&\quad-\sum_{k'=3}^n\frac{1}{k'(k'-1)}\frac{{n-i-j-1\choose
k'-3}}{{n-1\choose k'-1}}\\
&=\frac{1}{i}\sum_{k'=3}^n\sum_{t=1}^{k'-2}\frac{1}{k'-1}\frac{{i\choose t}{n-i-j-1\choose
k'-t-2}}{{n-1\choose k'-1}}-\sum_{k'=3}^n\sum_{t=1}^{k'-2}\frac{1}{k'(k'-1)}\frac{{i-1\choose t-1}{n-i-j-1\choose
k'-t-2}}{{n-1\choose k'-1}}\\
&\quad-\sum_{k'=3}^n\frac{1}{k'(k'-1)}\frac{{n-i-j-1\choose
k'-3}}{{n-1\choose k'-1}}\\
&=\frac{1}{i}\sum_{k'=3}^n\left(\sum_{t=0}^{k'-2}\frac{1}{k'-1}\frac{{i\choose t}{n-j-1-i\choose k'-2-t}}{{n-1\choose
k'-1}}-\frac{1}{k'-1}\frac{{n-i-j-1\choose
k'-2}}{{n-1\choose k'-1}}\right)\\
&\quad-\sum_{k'=3}^n\sum_{t=0}^{k'-3}\frac{1}{k'(k'-1)}\frac{{i-1\choose t}{n-j-2-(i-1)\choose
k'-3-t}}{{n-1\choose k'-1}}-\sum_{k'=3}^n\frac{1}{k'(k'-1)}\frac{{n-i-j-1\choose
k'-3}}{{n-1\choose k'-1}}\\
&\overset{\mathclap{(\ref{eq:bi2})}}=\frac{1}{i}\sum_{k'=3}^n\left(\frac{1}{k'-1}\frac{{n-j-1\choose k'-2}}{{n-1\choose
k'-1}}-\frac{1}{k'-1}\frac{{n-i-j-1\choose
k'-2}}{{n-1\choose k'-1}}\right)\\
&\quad-\left(\sum_{k=3}^n\frac{1}{k'(k'-1)}\frac{{n-j-2\choose k'-3}}{{n-1\choose
k'-1}}+\sum_{k'=3}^n\frac{1}{k'(k'-1)}\frac{{n-i-j-1\choose k'-3}}{{n-1\choose k'-1}}\right)\\
&\overset{\mathclap{(14),(36)}}=\;\;\frac{1}{i}\sum_{k'=2}^n\left(\frac{{n-k'\choose j-1}}{{n-1\choose
j}}\frac{1}{j}-\frac{{n-k'\choose i+j-1}}{{n-1\choose i+j}}\frac{1}{i+j}\right)\\
&\quad\quad-\frac{1}{2}\left(\beta_n(j)-\beta_n(j+1)+\beta_n(i+j-1)-\beta_n(i+j)\right)\\
&\overset{\mathclap{(\ref{eq:bi1})}}=\frac{1}{ij}-\frac{1}{i(i+j)}-\frac{1}{2}\left(\beta_n(j)-\beta_n(j+1)+\beta_n(i+j-1)-\beta_n(i+j)\right)\\[4ex]
\intertext{Case $i+j=n$ ($\Rightarrow$ $k=2$ and $t=k'-1$)}
s_b(n-j,j)&=\sum_{k'=3}^nk'p_b(2,n-j;k',j)\frac{1}{k'(k'-1)}\\
&=\sum_{k'=3}^n\frac{1}{k'-1}\frac{1}{k'}\frac{{n-j-1\choose
t-1}}{{n-1\choose k'-1}}\frac{1}{k'-1}\\
&=\sum_{k'=3}^n\frac{{n-j-1\choose
t-1}}{{n-1\choose k'-1}}\frac{1}{k'(k'-1)^2}\\
&\overset{\mathclap{(14)}}=\sum_{k'=3}^n\frac{{n-k'\choose
j-1}}{{n-1\choose j}j}\frac{1}{k'(k'-1)}\\
&=\sum_{k'=2}^n\frac{{n-k'\choose j-1}}{{n-1\choose j}j}\frac{1}{k'(k'-1)}-\frac{{n-2\choose
j-1}}{{n-1\choose j}}\frac{1}{2j}\\
&=\sum_{k'=2}^n\frac{{n-k'\choose j-1}}{{n-1\choose
j}j}\left(\frac{1}{k'-1}-\frac{1}{k'}\right)-\frac{1}{2(n-1)}\\
&\overset{\mathclap{(\ref{eq:34a}),(34)}}=\quad\frac{a_n-a_j}{n-j}-\frac{1}{2}\beta_n(j)-\frac{1}{2(n-1)}
\end{align*}
\endgroup

\subsection{Derivation of the population spectrum}

The population 1-SFS spectrum of linked sites $\ev[\xi(f|f_0)]$ (equation \ref{sfs_ls}) can be derived from the 1-SFS sample spectrum $\ev[\xi_{k|l}]$ (equation \ref{sfs_lp}) by the formula
\beqs
\ev[\xi(f|f_0)]=\lim_{n\rightarrow\infty} n\ev[\xi_{\lfloor nf \rfloor|\lfloor nf_0 \rfloor}]
\eeqs
The derivation is a cumbersome but relatively simple computation, once we prove a few limits and asymptotic results. 

The ``big O'' notation $O(x_n)$ is used for a function of $n$ that behaves asymptotically as $x_n$ for $n\rightarrow\infty$, i.e. $O(x_n)/x_n\rightarrow$ constant. The indicator function $I(A)$ is 1 if $A$ is true and 0 otherwise.

First, we state two useful asymptotic results:
\beqs
\lfloor nf \rfloor=nf\cdot(1+O(1/n))
\eeqs
\beqs
a_n=\ln(n)+\gamma+O(1/n)=(\ln(n)+\gamma)\cdot(1+O(1/n\ln(n)))
\eeqs
where $\gamma$ is the Eulero-Mascheroni constant.

The main derivation involves the limits of a few terms. The first one is $nl(\beta_n(k)-\beta_n(k+1))$. By some manipulations:
\begin{align*}
&nl(\beta_n(k)-\beta_n(k+1))=\nonumber\\
&=\frac{-4n^2l(a_{n+1}-a_k)}{(n-k+1)(n-k)(n-k-1)}+\frac{2n^2l(a_{k+1}-a_k)}{(n-k+1)(n-k)}+\frac{2nl}{(n-k)(n-k-1)}=\nonumber\\
&=\nonumber\left[\frac{-4(l/n)(\ln(n+1)-\ln(k))}{(1-k/n+1/n)(1-k/n)(1-k/n-1/n)}+\frac{2(l/n)/(k/n)}{(1-k/n+1/n)(1-k/n)}+\right.\nonumber\\
&\left.+\frac{2l/n}{(1-k/n)(1-k/n-1/n)}\right]\cdot(1+O(1/n))\nonumber
\end{align*}
hence
\beqs
n\lfloor nf_0\rfloor(\beta_n(\lfloor nf\rfloor)-\beta_n(\lfloor nf\rfloor+1))\ \mathop{\longrightarrow}_{n\rightarrow\infty}\ \frac{4f_0\ln(f)}{(1-f)^3}+\frac{2f_0/f}{(1-f)^2}+\frac{2f_0}{(1-f)^2}
\eeqs
The second one is $l\frac{a_n-a_k}{n-k}$:
\beqs
l\frac{a_n-a_k}{n-k}=\frac{l}{n}\frac{\ln(n)-\ln(k)}{1-k/n}\cdot(1+O(1/n))
\eeqs
hence
\beqs
\lfloor nf_0\rfloor\frac{a_n-a_\lfloor nf\rfloor}{n-\lfloor nf\rfloor}\ \mathop{\longrightarrow}_{n\rightarrow\infty}\ f_0\frac{-\ln(f)}{1-f}
\eeqs
The last one is $l\beta_n(k)$:
\beqs
l\beta_n(k)=\frac{l}{n}\left(\frac{\ln(n)-\ln(k)}{(1-k/n)(1-k/n-1/n)}-\frac{2}{1-k/n}\right)\cdot (1+O(1/n))
\eeqs
hence
\beqs
\lfloor nf_0\rfloor\beta_n(\lfloor nf\rfloor)\ \mathop{\longrightarrow}_{n\rightarrow\infty}\ f_0\left(\frac{-2\ln(f)}{(1-f)^2}-\frac{2}{1-f}\right)
\eeqs

Finally, the limit of the Kronecker delta $\delta_{k,l}$ (which appears implicitly as a multiplicative factor in the spectrum for co-occurring and complementary mutations) is a non-trivial one. In fact, the limit 
\beqs
n\delta_{\lfloor nf\rfloor ,\lfloor nf_0\rfloor }\rightarrow \delta(f-f_0)
\eeqs
exists only as a convergence in the space of distributions. We prove it directly by showing that the two distributions converge when applied to an arbitrary smooth test function $h(f)$ with compact support:
\begin{align*}
&\int_{-\infty}^\infty df\ h(f)\left[n\delta_{\lfloor nf\rfloor ,\lfloor nf_0\rfloor }- \delta(f-f_0)\right]=\nonumber\\
&=n\int_{-\infty}^\infty df\ h(f)\ I(\lfloor nf_0\rfloor \leq nf < \lfloor nf_0\rfloor+1) -h(f_0)=\qquad (\mathrm{use}\ x=nf-\lfloor nf_0\rfloor)\nonumber\\
&=n\int_{0}^1 dx\ h(x/n+\lfloor nf_0\rfloor/n) -h(f_0)\ \mathop{\longrightarrow}_{n\rightarrow\infty}\ h(f_0)-h(f_0)=0\nonumber\\
\end{align*}

\subsection{Derivation of the triallelic spectrum}

The mutation process for two non-recombining loci - resulting in the generation of three alleles - resembles the mutation process for a single triallelic locus once the different mutation rates are taken into account. The rescaled mutation rates for the two mutations are ($\theta^{loc}_1$, $\theta^{loc}_2$) instead of ($2\theta$, $\theta$) for the two-site case (the first mutation can appear in either of the loci, hence the factor of 2). Moreover, we consider a single locus instead of $L(L-1)/2\sim L^2/2$ pairs of sites. The overall factor is therefore $\theta^{loc}_1\theta^{loc}_2/\theta^2L^2$. Both nested and disjoint components contribute to the triallelic spectrum.

\end{document}